\documentclass[runningheads]{llncs}
\usepackage[T1]{fontenc}
\usepackage{graphicx}
\usepackage{booktabs}
\usepackage{textcomp}
\usepackage{xcolor}
\usepackage{subfig}
\usepackage{array}
\usepackage{multirow}

\begin{document}
\title{Beyond Functional Correctness: Static Code Quality Profiling of LLM-Generated Code}
\titlerunning{Static Code Quality Profiling of LLM-Generated Code}
%
%
\author{Jue Huang\inst{1} \and
Tarek Mahmud\inst{2} \and
Meiru Che\inst{3} \and
Guowei Yang\inst{1}}
\authorrunning{J. Huang et al.}
%
\institute{
The University of Queensland, Brisbane, QLD, Australia\\
\email{{jue.huang, guowei.yang}@uq.edu.au}
\and
Texas A\&M University--Kingsville, Kingsville, TX, USA\\
\email{Tarek.Mahmud@tamuk.edu}
\and
Central Queensland University, Brisbane, QLD, Australia\\
\email{m.che@cqu.edu.au}
}
\maketitle              
\begin{abstract}
Large language models (LLMs) are increasingly used for code generation, yet most evaluations still focus on whether generated code passes benchmark test suites. Functional correctness alone, however, provides only a limited view of quality: benchmark tests are often incomplete, and passing them does not guarantee that code is maintainable, reliable, or well-structured. This paper {presents a large-scale static quality profiling of} LLM-generated Python code, with a focus on maintainability and reliability. Using 65{,}663 programs from 73 models in LiveCodeBench, we evaluate generated code with established static analysis tools, including Pylint, Radon, and SonarQube. Our results show systematic differences across models that are not captured by correctness alone, with maintainability emerging as the clearest source of variation. We further find that correctness often comes with structural trade-offs rather than uniformly simpler solutions. Overall, our findings motivate evaluation frameworks that assess LLM-generated code beyond pass rates alone.

\keywords{Large Language Models \and Code Generation \and Static Analysis
\and {Code Quality Profiling} \and Maintainability}
\end{abstract}
\section{Introduction}

LLMs are increasingly used in software development to assist with programming tasks, generate candidate solutions, and support rapid prototyping. As instruction-tuned and code-specialized models become more common, automated code generation is becoming routine in both industry and academia. This makes it increasingly important to understand not only whether generated code works, but also its structural and non-functional properties.

To the best of our knowledge, recent evaluations focus on functional correctness, typically measured by whether generated programs pass benchmark test suites such as HumanEval~\cite{chen2021evaluating} and MBPP~\cite{austin2021program}. While this has been useful for establishing common baselines, correctness alone provides only a limited view of code quality. A program may pass benchmark tests yet still be unnecessarily complex, hard to read, or difficult to maintain. 

Prior work suggests that LLM-generated code can exhibit maintainability and structural quality issues, including unnecessary complexity, naming problems, and other code-quality concerns~\cite{idrisov2024program,11185875}.

Software engineering research has long emphasized the importance of non-functional attributes such as maintainability, reliability, and security. As LLM-generated code becomes part of real development workflows, these non-functional dimensions deserve systematic evaluation.

Despite growing interest in LLM-based code generation, the structural properties of generated code remain underexplored at scale. Existing benchmarks provide limited visibility into non-functional quality, and relatively few studies~\cite{11126649,11185875} evaluate generated code using static analysis tools commonly employed in practice. Afrin et al.~\cite{11185875} examine how quantization affects the functional performance and static code-quality properties of large code-generation models. However, it remains unclear what static code-quality profiles different LLMs exhibit, which quality dimensions most clearly distinguish them, and how these profiles vary across task difficulty levels.

To address this gap, we present a large-scale static code-quality profiling study of LLM-generated Python programs using a broad set of static quality metrics. We combine widely used software engineering tools—Pylint~\cite{Pylint_contributors_Pylint}, Radon~\cite{lacchia2023radon}, and SonarQube~\cite{papapetrou2013sonarqube}—to analyze code style, refactoring signals, structural complexity, maintainability, reliability, and security-related warnings. Our study covers 65{,}663 programs generated by 73 LLMs from the LiveCodeBench~\cite{jain2024livecodebench} dataset, spanning diverse model families and task difficulty levels. We use these complementary metrics to characterize and compare multidimensional static quality profiles across models and across Easy, Medium, and Hard tasks.

To support the interpretation of these profiles, we additionally analyze functionally correct programs separately and report length-normalized metrics for the correct-only subset. These analyses help determine whether the observed patterns are influenced by incomplete failed implementations or differences in code volume. We treat them as supporting analyses rather than separate research objectives.

Our analysis reveals systematic differences in how LLMs structure code, identifying patterns that are not captured by correctness-based evaluation alone. In particular, maintainability emerges as a key dimension for distinguishing model behavior. As task difficulty increases, generated programs generally become longer and more structurally involved, while the corresponding changes in static-analysis findings depend on whether raw counts or length-normalized measures are considered.

This work makes the following contributions toward profiling the static quality of LLM-generated code beyond functional correctness:

\begin{itemize}

\item \textbf{A large-scale static code-quality profiling study of LLM-generated code.}
Using established analysis tools—Pylint, Radon, and SonarQube—we profile maintainability, reliability, security-related findings, structural complexity, and code size in 65{,}663 Python programs generated by 73 LLMs in the LiveCodeBench dataset.

\item \textbf{Characterization of cross-model and cross-difficulty code-quality profiles.}
We identify consistent stylistic and structural tendencies across LLMs and show how multiple static quality dimensions vary across Easy, Medium, and Hard programming tasks.

\item \textbf{Reproducibility through artifact availability.}
We release all code, analysis scripts, and aggregated experimental results to support reproducibility and enable further investigation\footnote{\url{https://github.com/JueH-96/Code-Quality-Evaluation.git}}.

\end{itemize}

\section{Background}

\subsection{LLMs for Software Engineering and Code Generation}

LLMs such as ChatGPT, Gemini, and Claude are increasingly used for code generation~\cite{jiang2024survey,nejjar2025llms,liu2024exploring,joel2024survey}, code repair~\cite{joshi2023repair,yang2024cref,bouzenia2024repairagent}, code translation~\cite{zhang2023planning,pan2024lost,tao2024unraveling}, and code completion~\cite{husein2025large,cheng2025security}. Although they perform well on many benchmarks, they continue to struggle with complex programming tasks~\cite{rao2025navigating,mohsin2024can,zhong2024can,dou2024s}, motivating research on improving the reliability and functional correctness of generated code~\cite{mu2024clarifygpt,dong2024self,nijkamp2022codegen,gao2025current}.

The challenges of LLM-based code generation extend beyond syntactic validity. Generated solutions may omit required logic, mishandle edge cases, introduce unnecessary control flow, or terminate before completing the implementation~\cite{dou2024s,mohsin2024can,zhong2024can}. Consequently, short or structurally simple code should not automatically be interpreted as high-quality code, particularly when the implementation is incomplete.

\subsection{Limits of Correctness-Based Evaluation}

Benchmarks such as HumanEval~\cite{chen2021evaluating}, MBPP~\cite{austin2021program}, and DS-1000~\cite{lai2023ds} primarily evaluate functional correctness using predefined test suites. While this approach enables standardized model comparison, its validity depends on the quality and coverage of the tests. A program may pass the available tests while still failing on untested inputs or containing brittle logic. EvalPlus~\cite{liu2023your}, for example, shows that stronger mutation-augmented tests can substantially change evaluation outcomes. Passing a limited test suite also does not establish that generated code is maintainable, robust, or free from structural and reliability-related issues~\cite{chou2001empirical,palix2011faults}.

Constructing comprehensive test suites is difficult and costly~\cite{brunet2009design,muccini2004using,aniche2022effective,heumann2001generating}. Dynamic approaches such as fuzzing can improve coverage~\cite{aschermann2019redqueen,xia2024fuzz4all}, but they are computationally expensive and difficult to apply consistently at scale. Static analysis therefore provides a complementary means of examining code characteristics beyond benchmark pass rates.

Correctness status must also be considered when interpreting static metrics. Incorrect solutions may appear shorter or less complex because they omit required operations, branches, or entire parts of the implementation. Analyses that combine correct and incorrect solutions may therefore conflate implementation completeness with non-functional quality. Examining correct solutions separately provides a more controlled comparison among programs that satisfy the same benchmark-based correctness criterion.

\subsection{Static Analysis and Code Quality Metrics}

Maintainability and reliability are central dimensions of long-term software quality~\cite{baggen2012standardized,syed2015guidelines,braz2022software}. Poor code quality can consume substantial engineering effort, with prior work estimating that 23--42\% of development time is spent addressing quality-related problems~\cite{tornhill2022code}. Industry reports similarly identify technical debt and accumulated quality deficiencies as major drains on development resources~\cite{Sonar2023,McKinsey2023}.

Static-analysis tools such as Pylint, Radon, and SonarQube are commonly used to identify coding-rule violations, refactoring opportunities, structural complexity, maintainability issues, and potential reliability problems~\cite{copeland2005pmd,Pylint_contributors_Pylint,lacchia2023radon,papapetrou2013sonarqube}. Static analysis is suitable for large-scale evaluation because it does not require executable test cases and can be applied consistently across large code corpora. Dynamic methods, in contrast, often face scalability challenges such as path explosion~\cite{krishnamoorthy2010tackling}.

However, static-analysis results must be interpreted in relation to program size and completeness. Longer programs naturally provide more opportunities for tools to report issues, so raw counts may reflect both code quality and verbosity. Length-normalized measures, such as issues per 100 source lines of code, help distinguish total issue burden from issue concentration. Similarly, low code size or cyclomatic complexity may indicate concise code, but may also result from incomplete implementations. These measures should therefore be treated as structural characteristics rather than inherently positive or negative indicators of quality.

Existing evaluations of LLM-generated code continue to focus primarily on functional correctness~\cite{chen2021evaluating,austin2021program,lai2023ds,liu2023your}, with less attention to large-scale comparisons of static code quality across models. Therefore, we investigate the static code quality profile of LLM-generated code. 
In this study, a static code quality profile is the vector of static-analysis and structural measures associated with the code generated by an LLM, including coding-rule findings, complexity, code size, documentation-related characteristics, and maintainability-, reliability-, and se\-curity-related tool reports. A profile describes recurring characteristics of generated code and is not intended as an overall ranking of model quality.

\section{Methodology}

\subsection{Research Questions}

This study investigates the static code quality of LLM-generated code. We focus on {two} research questions:

\begin{itemize}
    \item \textbf{RQ1:} How do static code quality metrics vary across different LLMs?
    \item \textbf{RQ2:} What trade-offs emerge between functional correctness and static code quality metrics in LLM-generated code?
\end{itemize}

RQ1 is motivated by the fact that LLMs may achieve similar correctness while differing in maintainability, reliability, and security-related quality. RQ2 is motivated by the fact that passing benchmark tests does not guarantee high-quality code, and higher correctness may come with trade-offs such as greater complexity or lower maintainability.  To answer these questions, we build a dataset of Python programs generated by a broad set of LLMs, evaluate these programs using established static analysis tools, and analyze the resulting metrics.

\subsection{Experimental Setup}

\subsubsection{Dataset}\label{AA}

We use LiveCodeBench as the source of LLM-generated code in this study and collect all available program submissions from the public LiveCodeBench submissions repository\footnote{\url{https://github.com/LiveCodeBench/submissions}}.

The benchmark contains submissions from up to 73 LLMs across up to 1{,}055 prompts, enabling large-scale analysis. Each prompt includes 10 generated code samples to support pass@10 evaluation. For consistency, we select the first code sample for each prompt and model, which is typically closest to the model's highest-ranked completion~\cite{chen2021evaluating}. After extraction and filtering, the final dataset contains 65{,}663 code samples.

Each prompt is labeled as Easy, Medium, or Hard in LiveCodeBench. We preserve this partitioning and analyze code quality metrics separately for each difficulty level.

We conduct the analysis using two complementary subsets. The \emph{all-code} subset contains all 65{,}663 generated programs, including both correct and incorrect solutions, and represents the complete distribution of model outputs. The \emph{correct-only} subset contains only programs that pass the corresponding LiveCodeBench tests. This second subset reduces the influence of incorrect programs that may contain incomplete implementations, missing branches, or partially developed logic and therefore appear artificially short or structurally simple.

\subsubsection{Non-functional Qualities}

Our evaluation focuses on three non-functional software qualities: \emph{maintainability}, \emph{reliability}, and \emph{security}. We also consider functional correctness as reported by LiveCodeBench, but treat it as an external benchmark signal rather than a complete measure of code quality.

\textbf{Maintainability}. Maintainability reflects how easily code can be understood, modified, and extended. We evaluate it using metrics related to complexity, coding conventions, refactoring signals, comment usage, and maintainability-related issues reported by static analysis tools.

\textbf{Reliability}. Reliability concerns the likelihood that code executes without failure under expected conditions. We assess it using static indicators such as warnings, potential errors, fatal errors, and reliability-related issues identified by the analysis tools.

\textbf{Security}. Security concerns vulnerabilities and risky coding patterns that could affect confidentiality, integrity, or availability. We evaluate it using security-focused static analysis reports that identify potentially unsafe code patterns.

\subsubsection{Quality Construct Operationalization}

We operationalize the selected quality constructs using complementary static-analysis indicators rather than a single composite score. Maintainability is represented by Pylint convention/refactor findings, Radon complexity and Maintainability Index, and SonarQube maintainability issues; reliability by Pylint warning/error findings and SonarQube reliability issues; and security by SonarQube security findings. Code-size and organizational measures are treated as structural context rather than direct quality indicators. We retain the individual metrics to avoid subjective weighting across measures with different meanings and scales.

\subsubsection{Tools Selection}

We use three code evaluation tools to measure maintainability, reliability, and security in Python code: Pylint (3.3.5), Radon (6.0.1), and SonarQube Community Edition (10.4). These tools are widely used in practice and provide complementary metrics covering style, structure, reliability, and security.

\textbf{Pylint}. Pylint is a static analysis tool that detects code quality issues before execution. We use four categories of findings: \textit{Convention} (style violations), \textit{Refactor} (signs of unnecessarily complex structure), \textit{Warning} (questionable or unintended usage), and \textit{Error} (definite coding mistakes that may lead to runtime failures). For each category, we count detected issues and compute the average per code sample. We include Pylint because it is one of the most widely used Python analyzers and captures a broad range of maintainability- and reliability-related signals beyond style checking alone~\cite{gulabovska2019survey}.

\textbf{Radon}. Radon reports structural properties related to complexity and maintainability. We use cyclomatic complexity, maintainability index, function count, LOC, SLOC, LLOC, and counts of comment and blank lines. Maintainability index is a score-based metric, while the others capture structural properties of the code. For each metric, we compute the average across code samples. We include Radon because it is commonly used to measure complexity and maintainability in Python and is well suited to workflow integration~\cite{hassan2024evaluating}.

\textbf{SonarQube}. SonarQube is a code analysis platform designed to assess code quality and security at scale. In our study, we use three categories of findings: \textit{Maintainability Issues}, \textit{Reliability Issues}, and \textit{Security Issues}. All SonarQube metrics are expressed as counts, and we compute the average number of issues per code sample. We include SonarQube because prior large-scale evidence suggests that its findings capture meaningful software-quality signals, even if imperfectly~\cite{lenarduzzi2020some}.

\subsubsection{{Length-Normalized Metrics}}

{As raw static-analysis counts increase with program length, we normalize count-based metrics by lines of code. For each model or model--difficulty group (g), the density of metric (m) is:}

\begin{equation}
\mathrm{Density}_{g,m}
=
\frac{\sum_{i \in g} \mathrm{Count}_{i,m}}
{\sum_{i \in g} \mathrm{SLOC}_{i}}
\times 100.
\end{equation}

{This value represents findings per 100 SLOC. We apply normalization to count-based Pylint and SonarQube metrics and, where appropriate, to structural measures such as complexity, comments, blank lines, and function counts. Code-size metrics and the Maintainability Index remain in their original forms.}

\subsubsection{Experiments}

We extracted 65{,}663 code samples and applied the three analysis tools to collect all metrics. For each sample, we record its generating model, task difficulty, LiveCodeBench correctness label, and static-analysis results.

To answer \textbf{RQ1}, we compare raw metric distributions across LLMs and across Easy, Medium, and Hard tasks. We repeat the analysis on functionally correct programs and use length-normalized metrics to assess whether the observed patterns persist after excluding potentially incomplete implementations and accounting for code volume.

To answer \textbf{RQ2}, we compare the static quality metrics of correct and incorrect programs across difficulty levels. We further analyze raw and length-normalized metrics for correct programs to distinguish effects associated with implementation completeness from those related to code length.

\section{Experimental Results}

\subsection{RQ1: How do static code quality metrics vary across different LLMs?}

\begin{table}[t!]
\centering
\caption{
Summary of code quality metrics across Easy, Medium, and Hard tasks for the best-performing LLM from each organisation, selected based on the highest functional correctness rate.
}
\label{tab:code_quality_best_per_org_all_difficulties}
\setlength{\tabcolsep}{2pt}
\renewcommand{\arraystretch}{1.12}
\fontsize{6.1}{7.0}\selectfont

\resizebox{\linewidth}{!}{%
\begin{tabular}{llcccccccccccccccc}
\toprule
\textbf{LLM} & \textbf{Diff.}
& \multicolumn{5}{c}{\textbf{Pylint}}
& \multicolumn{8}{c}{\textbf{Radon}}
& \multicolumn{3}{c}{\textbf{SonarQube}} \\
\cmidrule(lr){3-7}
\cmidrule(lr){8-15}
\cmidrule(lr){16-18}
&
& \textbf{Issues} & \textbf{Conv} & \textbf{Ref} & \textbf{Warn} & \textbf{Err}
& \textbf{Comp} & \textbf{MI} & \textbf{LOC} & \textbf{LLOC} & \textbf{SLOC} & \textbf{CMT} & \textbf{BLN} & \textbf{Func}
& \textbf{Maint} & \textbf{Rel} & \textbf{Sec} \\
\midrule

\multirow{3}{*}{O3 (High)}
& Easy   &  6.40 &  5.81 & 0.46 &  0.07 & 0.06 &  6.29 & 87.34 & 21.46 & 12.72 & 12.30 &  3.61 &  3.94 & 1.40 & 0.69 & 0.02 & 0.00 \\
& Medium &  7.51 &  6.39 & 0.87 &  0.19 & 0.07 & 11.45 & 79.26 & 39.29 & 22.87 & 22.25 &  7.44 &  6.59 & 1.63 & 1.22 & 0.01 & 0.00 \\
& Hard   &  9.55 &  7.82 & 1.23 &  0.33 & 0.18 & 15.92 & 71.18 & 66.36 & 37.64 & 36.64 & 11.35 &  9.88 & 1.68 & 2.18 & 0.01 & 0.00 \\
\midrule

\multirow{3}{*}{Kimi-k1.6-IOI-high}
& Easy   &  4.75 &  3.93 & 0.53 &  0.02 & 0.27 &  8.94 & 73.80 & 11.64 & 10.31 & 10.16 &  0.59 &  0.94 & 0.87 & 0.51 & 0.12 & 0.00 \\
& Medium &  9.88 &  8.11 & 1.08 &  0.26 & 0.44 & 12.87 & 64.50 & 29.69 & 24.34 & 24.07 &  1.90 &  3.66 & 1.46 & 1.36 & 0.02 & 0.00 \\
& Hard   & 15.66 & 12.83 & 1.67 &  0.81 & 0.35 & 15.79 & 57.29 & 56.30 & 44.23 & 43.64 &  4.21 &  7.01 & 1.61 & 2.36 & 0.02 & 0.00 \\
\midrule

\multirow{3}{*}{DeepSeek-R1-0528}
& Easy   &  6.38 &  5.39 & 0.59 &  0.07 & 0.34 &  6.50 & 65.12 & 14.39 & 12.91 & 12.80 &  0.02 &  1.55 & 1.35 & 0.62 & 0.29 & 0.00 \\
& Medium &  9.67 &  7.83 & 1.21 &  0.11 & 0.52 & 13.09 & 54.18 & 29.44 & 25.26 & 24.99 &  0.01 &  3.60 & 1.55 & 1.28 & 0.03 & 0.00 \\
& Hard   & 14.35 & 11.59 & 2.03 &  0.35 & 0.38 & 20.56 & 43.36 & 60.87 & 48.59 & 47.83 &  0.04 &  6.76 & 1.64 & 2.34 & 0.06 & 0.00 \\
\midrule

\multirow{3}{*}{Gemini-2.5-Pro-06-05}
& Easy   &  9.22 &  6.06 & 0.48 &  2.63 & 0.05 &  6.17 & 81.85 & 34.76 & 13.57 & 12.64 & 11.93 &  5.90 & 1.34 & 0.63 & 0.05 & 0.00 \\
& Medium & 12.47 &  9.95 & 0.84 &  1.58 & 0.10 & 11.22 & 77.63 & 56.88 & 25.31 & 24.28 & 16.62 & 10.21 & 1.54 & 1.43 & 0.01 & 0.00 \\
& Hard   & 18.22 & 14.44 & 1.64 &  2.02 & 0.11 & 16.49 & 70.03 & 89.06 & 48.51 & 46.25 & 17.42 & 15.09 & 1.57 & 2.64 & 0.03 & 0.00 \\
\midrule

\multirow{3}{*}{Grok-3-Mini (High)}
& Easy   &  4.99 &  4.17 & 0.53 &  0.02 & 0.26 &  8.93 & 81.38 & 15.03 & 11.18 & 11.01 &  2.72 &  1.78 & 0.84 & 0.49 & 0.05 & 0.00 \\
& Medium &  8.71 &  7.27 & 1.07 &  0.19 & 0.19 & 13.54 & 72.87 & 34.23 & 25.68 & 25.47 &  5.93 &  4.15 & 1.24 & 1.22 & 0.01 & 0.00 \\
& Hard   & 11.71 &  9.42 & 1.18 &  1.01 & 0.10 & 15.46 & 66.48 & 59.57 & 45.84 & 45.32 &  9.59 &  6.83 & 1.30 & 1.70 & 0.01 & 0.00 \\
\midrule

\multirow{3}{*}{OpenReasoning-Nemotron-32B}
& Easy   & 16.38 &  5.02 & 0.56 & 10.48 & 0.31 &  7.56 & 65.57 & 14.22 & 12.70 & 12.55 &  0.03 &  1.47 & 1.21 & 0.65 & 0.22 & 0.00 \\
& Medium & 35.03 &  8.52 & 1.27 & 24.84 & 0.40 & 14.03 & 52.99 & 32.51 & 27.45 & 27.15 &  0.04 &  4.24 & 1.53 & 1.47 & 0.02 & 0.00 \\
& Hard   & 61.92 & 12.28 & 2.22 & 47.05 & 0.37 & 20.99 & 42.41 & 63.33 & 50.43 & 49.66 &  0.00 &  7.45 & 1.63 & 2.19 & 0.03 & 0.00 \\
\midrule

\multirow{3}{*}{EXAONE-4.0-32B}
& Easy   &  6.70 &  5.76 & 0.57 &  0.07 & 0.29 &  6.53 & 64.19 & 15.81 & 13.57 & 13.48 &  0.02 &  1.93 & 1.36 & 0.66 & 0.17 & 0.00 \\
& Medium & 11.10 &  9.26 & 1.25 &  0.13 & 0.45 & 13.84 & 52.98 & 33.57 & 27.73 & 27.46 &  0.04 &  4.69 & 1.59 & 1.53 & 0.02 & 0.00 \\
& Hard   & 15.09 & 12.49 & 1.98 &  0.23 & 0.40 & 19.99 & 42.95 & 61.46 & 46.73 & 46.21 &  0.01 &  7.47 & 1.47 & 2.33 & 0.01 & 0.00 \\
\midrule

\multirow{3}{*}{Qwen3-235B-A22B}
& Easy   &  5.18 &  4.51 & 0.53 &  0.09 & 0.05 &  8.10 & 75.71 & 14.31 & 11.56 & 11.36 &  1.15 &  1.76 & 0.99 & 0.63 & 0.11 & 0.00 \\
& Medium & 10.50 &  8.71 & 1.20 &  0.41 & 0.17 & 13.15 & 65.82 & 34.43 & 27.06 & 26.74 &  2.66 &  4.58 & 1.56 & 1.59 & 0.01 & 0.00 \\
& Hard   & 15.30 & 12.18 & 1.70 &  1.15 & 0.26 & 17.11 & 59.32 & 63.28 & 44.61 & 43.87 &  6.12 &  7.97 & 1.48 & 2.65 & 0.01 & 0.00 \\
\midrule

\multirow{3}{*}{Claude-Opus-4 (Thinking)}
& Easy   &  5.70 &  4.89 & 0.45 &  0.01 & 0.35 &  8.16 & 89.20 & 14.78 &  9.98 &  9.93 &  2.79 &  2.33 & 0.86 & 0.49 & 0.20 & 0.00 \\
& Medium &  9.87 &  8.35 & 0.72 &  0.16 & 0.64 & 12.02 & 81.56 & 28.65 & 18.93 & 18.86 &  4.99 &  5.29 & 1.25 & 0.84 & 0.14 & 0.00 \\
& Hard   & 13.14 & 11.14 & 0.73 &  0.87 & 0.40 & 13.05 & 74.63 & 44.85 & 30.35 & 30.16 &  6.70 &  8.53 & 1.17 & 1.01 & 0.23 & 0.00 \\
\midrule

\multirow{3}{*}{Mistral-Large}
& Easy   &  5.49 &  4.28 & 0.51 &  0.60 & 0.10 &  7.41 & 77.57 & 18.83 & 12.57 & 12.23 &  1.77 &  3.58 & 1.17 & 1.09 & 0.01 & 0.00 \\
& Medium &  7.24 &  5.63 & 0.72 &  0.84 & 0.06 &  9.55 & 71.86 & 28.17 & 19.66 & 19.21 &  2.53 &  5.26 & 1.59 & 1.62 & 0.01 & 0.00 \\
& Hard   &  7.96 &  6.07 & 0.52 &  1.24 & 0.12 & 10.14 & 64.98 & 39.01 & 26.24 & 25.72 &  2.47 &  6.83 & 1.45 & 1.84 & 0.01 & 0.00 \\
\bottomrule
\end{tabular}%
}

\vspace{1mm}

\footnotesize
\textbf{Note.}
All metrics are raw, unnormalized per-sample averages computed over both correct and incorrect generated programs.
For each organisation, the LLM with the highest functional correctness rate was selected as the representative model.

\textit{Pylint:}
Issues = total issues;
Conv = convention;
Ref = refactor;
Warn = warnings;
Err = errors.

\textit{Radon:}
Comp = cyclomatic complexity;
MI = maintainability index;
LOC = lines of code;
LLOC = logical LOC;
SLOC = source LOC;
CMT = comment lines;
BLN = blank lines;
Func = function count.

\textit{SonarQube:}
Maint = maintainability issues;
Rel = reliability issues;
Sec = security issues.
\end{table}

To answer \textbf{RQ1}, we construct and compare multidimensional static quality profiles across LLMs and task difficulty levels. These profiles combine Pylint findings, Radon structural measures, and SonarQube issues to characterize coding conventions, refactoring signals, structural complexity, code size, maintainability, reliability, and security-related findings. We first analyze raw metrics across all generated programs to capture the complete output profile of each model. We then use correct-only and length-normalized results as supporting analyses to assess whether the observed patterns are influenced by incomplete implementations or differences in code volume. Overall, maintainability-related and structural metrics provide the clearest differentiation among models, reliability-related findings vary more moderately, and security-related findings remain sparse. Table~\ref{tab:code_quality_best_per_org_all_difficulties} presents the raw profiles of representative LLMs.

\textit{Maintainability.}
The raw profiles reveal substantial cross-model variation in code structure. Pylint convention findings account for the largest share of reported issues for most models, while Radon shows corresponding differences in cyclomatic complexity, Maintainability Index, code length, comments, blank lines, and function counts. Some models generate relatively compact solutions with fewer findings, whereas others produce longer and more structurally involved implementations with higher raw issue counts. SonarQube maintainability findings show a similar pattern. These results represent distinct generation profiles rather than a simple quality ranking because raw counts are influenced by both code structure and code volume.

A similar shift appears across task difficulty levels. From Easy to Medium and Hard tasks, generated programs generally become longer and more structurally involved. Pylint issue counts increase, with convention findings remaining dominant and warning findings becoming more common. Radon metrics show increases in LOC, LLOC, SLOC, and cyclomatic complexity, accompanied by a decrease in Maintainability Index. SonarQube maintainability issue counts also tend to increase with difficulty. However, these raw changes do not establish that the concentration of findings per unit of code also increases.

\textit{Reliability.}
Reliability-related findings vary less than maintainability-related findings but still differ across models and difficulty levels. Pylint error counts remain low for most models, while warning findings contribute more strongly to cross-model variation. SonarQube reliability issue counts are also modest overall, indicating that reliability-related differences are less pronounced than maintainability-related differences.

Across difficulty levels, raw reliability-related counts increase more moderately than maintainability-related counts. Pylint warning findings generally increase with difficulty, whereas error findings remain comparatively rare. SonarQube reliability issue counts also tend to rise from Easy to Hard, although less sharply than maintainability issue counts. Length-normalized analysis is therefore needed to determine whether these increases reflect greater issue density or simply the larger volume of code generated for harder tasks.

\textit{Security.}
Security-related findings are rare across the evaluated models. This sparsity is likely influenced by the function-level nature of LiveCodeBench, which primarily contains short competitive-programming tasks with limited security-relevant functionality and attack surface. Consequently, the generated programs provide relatively few opportunities for static analysis tools to identify meaningful security issues.

\begin{table}[t!]
\centering
\caption{
Summary of raw code quality metrics for functionally correct programs across Easy, Medium, and Hard tasks. Results are shown for the same representative LLMs used in Table~\ref{tab:code_quality_best_per_org_all_difficulties}.
}
\label{tab:correct_only_quality_best_per_org}
\setlength{\tabcolsep}{2pt}
\renewcommand{\arraystretch}{1.12}
\fontsize{6.1}{7.0}\selectfont

\resizebox{\linewidth}{!}{%
\begin{tabular}{llcccccccccccccccc}
\toprule
\textbf{LLM} & \textbf{Diff.}
& \multicolumn{5}{c}{\textbf{Pylint}}
& \multicolumn{8}{c}{\textbf{Radon}}
& \multicolumn{3}{c}{\textbf{SonarQube}} \\
\cmidrule(lr){3-7}
\cmidrule(lr){8-15}
\cmidrule(lr){16-18}
&
& \textbf{Issues} & \textbf{Conv} & \textbf{Ref} & \textbf{Warn} & \textbf{Err}
& \textbf{Comp} & \textbf{MI} & \textbf{LOC} & \textbf{LLOC} & \textbf{SLOC}
& \textbf{CMT} & \textbf{BLN} & \textbf{Func}
& \textbf{Maint} & \textbf{Rel} & \textbf{Sec} \\
\midrule

\multirow{3}{*}{O3 (High)}
& Easy   & 6.41 & 5.81 & 0.47 & 0.07 & 0.06 & 5.94 & 82.12 & 20.16 & 12.70 & 12.29 & 3.59 & 3.93 & 1.40 & 0.69 & 0.02 & 0.00 \\
& Medium & 7.45 & 6.36 & 0.85 & 0.18 & 0.06 & 10.59 & 74.41 & 35.99 & 22.48 & 21.84 & 7.33 & 6.56 & 1.63 & 1.18 & 0.01 & 0.00 \\
& Hard   & 9.71 & 7.97 & 1.27 & 0.30 & 0.16 & 13.45 & 60.50 & 54.86 & 36.85 & 35.90 & 11.68 & 9.82 & 1.71 & 2.12 & 0.01 & 0.00 \\
\midrule

\multirow{3}{*}{Kimi-k1.6-IOI-high}
& Easy   & 4.74 & 3.92 & 0.53 & 0.01 & 0.27 & 4.02 & 73.51 & 11.53 & 10.27 & 10.12 & 0.58 & 0.92 & 0.87 & 0.51 & 0.12 & 0.00 \\
& Medium & 9.80 & 8.03 & 1.06 & 0.25 & 0.46 & 10.59 & 63.56 & 28.35 & 23.74 & 23.44 & 1.71 & 3.56 & 1.47 & 1.36 & 0.02 & 0.00 \\
& Hard   & 15.25 & 12.59 & 1.63 & 0.67 & 0.36 & 14.56 & 55.94 & 50.89 & 42.70 & 42.05 & 2.63 & 6.82 & 1.73 & 2.28 & 0.01 & 0.00 \\
\midrule

\multirow{3}{*}{DeepSeek-R1-0528}
& Easy   & 6.38 & 5.38 & 0.59 & 0.07 & 0.34 & 6.10 & 64.93 & 14.33 & 12.90 & 12.79 & 0.02 & 1.54 & 1.36 & 0.63 & 0.29 & 0.00 \\
& Medium & 9.49 & 7.72 & 1.13 & 0.11 & 0.53 & 11.46 & 53.23 & 27.44 & 24.16 & 23.91 & 0.01 & 3.54 & 1.52 & 1.21 & 0.03 & 0.00 \\
& Hard   & 13.40 & 11.00 & 1.70 & 0.25 & 0.45 & 15.50 & 40.70 & 48.60 & 42.76 & 42.08 & 0.03 & 6.50 & 1.65 & 1.85 & 0.03 & 0.00 \\
\midrule

\multirow{3}{*}{Gemini-2.5-Pro-06-05}
& Easy   & 9.25 & 6.07 & 0.49 & 2.65 & 0.05 & 5.65 & 81.31 & 34.54 & 13.57 & 12.64 & 11.94 & 5.91 & 1.34 & 0.63 & 0.05 & 0.00 \\
& Medium & 12.44 & 9.84 & 0.81 & 1.69 & 0.10 & 10.52 & 75.95 & 53.97 & 24.41 & 23.43 & 16.54 & 9.92 & 1.54 & 1.40 & 0.01 & 0.00 \\
& Hard   & 15.96 & 12.77 & 1.24 & 1.85 & 0.11 & 12.94 & 66.72 & 75.13 & 40.41 & 38.72 & 17.30 & 13.76 & 1.49 & 2.16 & 0.00 & 0.00 \\
\midrule

\multirow{3}{*}{Grok-3-Mini (High)}
& Easy   & 4.98 & 4.16 & 0.53 & 0.03 & 0.27 & 3.81 & 81.24 & 15.00 & 11.17 & 11.00 & 2.68 & 1.77 & 0.85 & 0.49 & 0.05 & 0.00 \\
& Medium & 8.77 & 7.25 & 1.12 & 0.19 & 0.21 & 8.98 & 72.35 & 32.95 & 24.94 & 24.72 & 5.67 & 4.05 & 1.30 & 1.24 & 0.01 & 0.00 \\
& Hard   & 12.49 & 10.09 & 1.17 & 1.16 & 0.07 & 10.55 & 65.39 & 60.25 & 46.27 & 45.87 & 9.96 & 7.28 & 1.42 & 1.81 & 0.01 & 0.00 \\
\midrule

\multirow{3}{*}{OpenReasoning-Nemotron-32B}
& Easy   & 16.00 & 4.96 & 0.55 & 10.17 & 0.32 & 5.79 & 64.92 & 13.62 & 12.38 & 12.19 & 0.03 & 1.40 & 1.21 & 0.65 & 0.22 & 0.00 \\
& Medium & 33.88 & 8.35 & 1.23 & 23.89 & 0.41 & 12.65 & 51.70 & 30.34 & 26.46 & 26.17 & 0.05 & 4.15 & 1.52 & 1.41 & 0.02 & 0.00 \\
& Hard   & 59.42 & 12.89 & 2.05 & 44.12 & 0.36 & 18.06 & 39.71 & 54.26 & 47.37 & 46.68 & 0.00 & 7.58 & 1.70 & 2.19 & 0.02 & 0.00 \\
\midrule

\multirow{3}{*}{EXAONE-4.0-32B}
& Easy   & 6.70 & 5.75 & 0.58 & 0.08 & 0.30 & 6.21 & 62.65 & 15.32 & 13.51 & 13.41 & 0.02 & 1.91 & 1.37 & 0.67 & 0.18 & 0.00 \\
& Medium & 10.83 & 9.02 & 1.22 & 0.12 & 0.47 & 12.66 & 51.27 & 30.63 & 26.46 & 26.17 & 0.05 & 4.45 & 1.62 & 1.45 & 0.01 & 0.00 \\
& Hard   & 14.98 & 12.46 & 1.86 & 0.23 & 0.44 & 16.74 & 40.05 & 50.89 & 44.29 & 43.61 & 0.02 & 7.27 & 1.62 & 2.15 & 0.01 & 0.00 \\
\midrule

\multirow{3}{*}{Qwen3-235B-A22B}
& Easy   & 5.19 & 4.51 & 0.53 & 0.09 & 0.05 & 4.64 & 75.17 & 14.20 & 11.55 & 11.36 & 1.14 & 1.75 & 1.00 & 0.63 & 0.10 & 0.00 \\
& Medium & 10.24 & 8.58 & 1.15 & 0.36 & 0.15 & 11.94 & 64.55 & 31.95 & 25.88 & 25.54 & 2.34 & 4.43 & 1.57 & 1.49 & 0.01 & 0.00 \\
& Hard   & 13.87 & 11.25 & 1.55 & 0.77 & 0.29 & 14.50 & 54.81 & 51.35 & 41.83 & 41.05 & 3.87 & 7.13 & 1.62 & 2.38 & 0.00 & 0.00 \\
\midrule

\multirow{3}{*}{Claude-Opus-4 (Thinking)}
& Easy   & 5.70 & 4.89 & 0.45 & 0.01 & 0.35 & 3.54 & 89.19 & 14.76 & 9.97 & 9.91 & 2.79 & 2.33 & 0.86 & 0.50 & 0.20 & 0.00 \\
& Medium & 9.77 & 8.28 & 0.72 & 0.12 & 0.64 & 7.95 & 82.18 & 27.46 & 18.18 & 18.14 & 4.75 & 5.06 & 1.31 & 0.86 & 0.13 & 0.00 \\
& Hard   & 12.36 & 10.58 & 0.66 & 0.62 & 0.51 & 7.10 & 75.96 & 41.11 & 27.39 & 27.28 & 6.41 & 8.12 & 1.16 & 0.85 & 0.12 & 0.00 \\
\midrule

\multirow{3}{*}{Mistral-Large}
& Easy   & 5.43 & 4.28 & 0.49 & 0.60 & 0.06 & 4.52 & 74.14 & 17.01 & 12.15 & 11.81 & 1.67 & 3.59 & 1.16 & 1.07 & 0.01 & 0.00 \\
& Medium & 6.80 & 5.30 & 0.66 & 0.79 & 0.06 & 7.91 & 67.42 & 25.29 & 18.57 & 18.09 & 2.28 & 5.00 & 1.53 & 1.51 & 0.00 & 0.00 \\
& Hard   & 10.23 & 7.38 & 0.46 & 2.31 & 0.08 & 7.23 & 52.23 & 42.77 & 33.15 & 32.31 & 1.31 & 9.31 & 1.46 & 2.77 & 0.00 & 0.00 \\
\bottomrule
\end{tabular}%
}

\vspace{1mm}

\footnotesize
\textbf{Note.}
All metrics are raw per-sample averages computed only over programs that pass the corresponding LiveCodeBench tests. The same representative model is used for each organisation as in Table~\ref{tab:code_quality_best_per_org_all_difficulties}.
\end{table}

\textit{Robustness of the profiles among correct-only programs.}
To assess whether the raw profiles are driven by incomplete failed implementations, we repeat the cross-model and cross-difficulty analyses using only programs that pass the LiveCodeBench tests. Table~\ref{tab:correct_only_quality_best_per_org} presents the resulting profiles for the same representative LLMs shown in Table~\ref{tab:code_quality_best_per_org_all_difficulties}.

The correct-only results preserve major cross-model differences in code size, structural complexity, coding conventions, Maintainability Index, and maintaina\-bility-related findings. Correct solutions to harder tasks also remain longer and more structurally involved than those for easier tasks. This suggests that the observed profiles are not explained solely by incomplete failed outputs, although removing incorrect programs changes the magnitude of some differences.

Reliability-related differences remain more moderate, while security-related findings remain sparse. However, because raw counts may still be affected by program length, we further examine length-normalized metrics.

\begin{table}[t!]
\centering
\caption{
Length-normalized code quality metrics for functionally correct programs across Easy, Medium, and Hard tasks. All values are reported per 100 SLOC for the same representative LLMs used in Table~\ref{tab:code_quality_best_per_org_all_difficulties}.
}
\label{tab:correct_only_per100sloc_best_per_org}
\setlength{\tabcolsep}{2.5pt}
\renewcommand{\arraystretch}{1.12}
\fontsize{6.3}{7.2}\selectfont

\resizebox{\linewidth}{!}{%
\begin{tabular}{llcccccccccccc}
\toprule
\textbf{LLM} & \textbf{Diff.}
& \multicolumn{5}{c}{\textbf{Pylint per 100 SLOC}}
& \multicolumn{4}{c}{\textbf{Radon per 100 SLOC}}
& \multicolumn{3}{c}{\textbf{SonarQube per 100 SLOC}} \\
\cmidrule(lr){3-7}
\cmidrule(lr){8-11}
\cmidrule(lr){12-14}
&
& \textbf{Issues} & \textbf{Conv} & \textbf{Ref} & \textbf{Warn} & \textbf{Err}
& \textbf{Comp} & \textbf{CMT} & \textbf{BLN} & \textbf{Func}
& \textbf{Maint} & \textbf{Rel} & \textbf{Sec} \\
\midrule

\multirow{3}{*}{O3 (High)}
& Easy   & 52.18 & 47.28 & 3.80 & 0.59 & 0.51 & 48.33 & 29.22 & 31.97 & 11.38 & 5.64 & 0.13 & 0.00 \\
& Medium & 34.12 & 29.11 & 3.89 & 0.83 & 0.29 & 48.48 & 33.59 & 30.06 & 7.48 & 5.42 & 0.03 & 0.00 \\
& Hard   & 27.04 & 22.19 & 3.55 & 0.84 & 0.46 & 37.47 & 32.54 & 27.36 & 4.75 & 5.91 & 0.02 & 0.00 \\
\midrule

\multirow{3}{*}{Kimi-k1.6-IOI-high}
& Easy   & 46.86 & 38.76 & 5.28 & 0.14 & 2.68 & 39.76 & 5.75 & 9.10 & 8.64 & 5.00 & 1.21 & 0.00 \\
& Medium & 41.82 & 34.25 & 4.54 & 1.08 & 1.95 & 45.17 & 7.28 & 15.19 & 6.28 & 5.80 & 0.08 & 0.00 \\
& Hard   & 36.27 & 29.93 & 3.88 & 1.60 & 0.84 & 34.63 & 6.25 & 16.21 & 4.11 & 5.43 & 0.03 & 0.00 \\
\midrule

\multirow{3}{*}{DeepSeek-R1-0528}
& Easy   & 49.88 & 42.08 & 4.62 & 0.54 & 2.64 & 47.73 & 0.12 & 12.07 & 10.61 & 4.91 & 2.30 & 0.00 \\
& Medium & 39.69 & 32.28 & 4.73 & 0.47 & 2.21 & 47.92 & 0.05 & 14.79 & 6.36 & 5.08 & 0.12 & 0.00 \\
& Hard   & 31.84 & 26.15 & 4.03 & 0.60 & 1.06 & 36.82 & 0.06 & 15.45 & 3.91 & 4.40 & 0.07 & 0.00 \\
\midrule

\multirow{3}{*}{Gemini-2.5-Pro-06-05}
& Easy   & 73.23 & 48.00 & 3.85 & 20.99 & 0.40 & 44.68 & 94.47 & 46.74 & 10.62 & 5.01 & 0.42 & 0.00 \\
& Medium & 53.08 & 41.97 & 3.46 & 7.22 & 0.44 & 44.89 & 70.58 & 42.32 & 6.59 & 5.98 & 0.02 & 0.00 \\
& Hard   & 41.22 & 32.97 & 3.20 & 4.77 & 0.27 & 33.43 & 44.68 & 35.53 & 3.86 & 5.58 & 0.01 & 0.00 \\
\midrule

\multirow{3}{*}{Grok-3-Mini (High)}
& Easy   & 45.25 & 37.77 & 4.83 & 0.23 & 2.42 & 34.64 & 24.37 & 16.11 & 7.71 & 4.43 & 0.43 & 0.00 \\
& Medium & 35.48 & 29.34 & 4.54 & 0.75 & 0.85 & 36.33 & 22.95 & 16.37 & 5.28 & 5.02 & 0.02 & 0.00 \\
& Hard   & 27.23 & 21.99 & 2.56 & 2.52 & 0.16 & 23.01 & 21.72 & 15.87 & 3.10 & 3.95 & 0.01 & 0.00 \\
\midrule

\multirow{3}{*}{OpenReasoning-Nemotron-32B}
& Easy   & 131.24 & 40.71 & 4.48 & 83.46 & 2.59 & 47.49 & 0.28 & 11.52 & 9.96 & 5.36 & 1.81 & 0.00 \\
& Medium & 129.45 & 31.90 & 4.71 & 91.28 & 1.57 & 48.32 & 0.18 & 15.87 & 5.79 & 5.37 & 0.07 & 0.00 \\
& Hard   & 127.28 & 27.61 & 4.39 & 94.52 & 0.77 & 38.69 & 0.00 & 16.23 & 3.64 & 4.68 & 0.04 & 0.00 \\
\midrule

\multirow{3}{*}{EXAONE-4.0-32B}
& Easy   & 49.94 & 42.86 & 4.32 & 0.56 & 2.21 & 46.33 & 0.16 & 14.22 & 10.23 & 5.02 & 1.31 & 0.00 \\
& Medium & 41.40 & 34.49 & 4.67 & 0.45 & 1.79 & 48.39 & 0.19 & 17.01 & 6.17 & 5.55 & 0.06 & 0.00 \\
& Hard   & 34.34 & 28.56 & 4.27 & 0.52 & 1.00 & 38.38 & 0.03 & 16.66 & 3.72 & 4.92 & 0.01 & 0.00 \\
\midrule

\multirow{3}{*}{Qwen3-235B-A22B}
& Easy   & 45.68 & 39.75 & 4.69 & 0.77 & 0.47 & 40.85 & 10.05 & 15.37 & 8.78 & 5.55 & 0.91 & 0.00 \\
& Medium & 40.08 & 33.59 & 4.49 & 1.42 & 0.59 & 46.74 & 9.15 & 17.34 & 6.16 & 5.83 & 0.02 & 0.00 \\
& Hard   & 33.78 & 27.42 & 3.78 & 1.88 & 0.71 & 35.33 & 9.42 & 17.36 & 3.96 & 5.80 & 0.00 & 0.00 \\
\midrule

\multirow{3}{*}{Claude-Opus-4 (Thinking)}
& Easy   & 57.56 & 49.37 & 4.59 & 0.06 & 3.54 & 35.76 & 28.20 & 23.54 & 8.70 & 5.00 & 2.03 & 0.00 \\
& Medium & 53.85 & 45.65 & 3.99 & 0.66 & 3.54 & 43.83 & 26.18 & 27.88 & 7.23 & 4.76 & 0.73 & 0.00 \\
& Hard   & 45.31 & 38.77 & 2.40 & 2.27 & 1.86 & 26.03 & 23.48 & 29.76 & 4.27 & 3.13 & 0.44 & 0.00 \\
\midrule

\multirow{3}{*}{Mistral-Large}
& Easy   & 45.99 & 36.21 & 4.18 & 5.12 & 0.49 & 38.31 & 14.15 & 30.37 & 9.82 & 9.07 & 0.08 & 0.00 \\
& Medium & 37.59 & 29.30 & 3.62 & 4.36 & 0.31 & 43.73 & 12.59 & 27.64 & 8.48 & 8.35 & 0.00 & 0.00 \\
& Hard   & 31.67 & 22.86 & 1.43 & 7.14 & 0.24 & 22.38 & 4.05 & 28.81 & 4.52 & 8.57 & 0.00 & 0.00 \\
\bottomrule
\end{tabular}%
}

\vspace{1mm}

\footnotesize
\textbf{Note.}
Values are calculated as the sum of each metric divided by the total SLOC within the corresponding model--difficulty group, multiplied by 100. Comp = cyclomatic complexity; CMT = comment lines; BLN = blank lines; Func = function count. The same representative model is used for each organisation as in Table~\ref{tab:code_quality_best_per_org_all_difficulties}. {Normalized structural measures represent their aggregate amount relative to code size and should not be interpreted as replacements for the corresponding raw structural metrics.}
\end{table}

\textit{Length normalized analysis.}
We next examine whether line of code impacts the number of issues. For this analysis, we normalize the issue count-based and selected structural metrics per 100 SLOC within the correct-only subset, as reported in Table~\ref{tab:correct_only_per100sloc_best_per_org}.

The normalized results show that higher issue counts on harder tasks do not always indicate higher issue density. For several models, raw Pylint and SonarQube findings increase from Easy to Hard, while the corresponding densities remain stable or decrease. For example, Qwen3-235B-A22B's total Pylint findings increase from 5.19 to 13.87, whereas its density decreases from 45.68 to 33.78 findings per 100 SLOC. Thus, part of the increase in raw findings is explained by the greater code volume required for harder tasks.

Cross-model differences nevertheless remain visible after normalization, particularly for convention-, warning-, and maintainability-related findings. OpenR\-easoning-Nemotron-32B, for instance, retains a high warning density across all difficulty levels, indicating that its raw warning counts are not attributable solely to longer code. Raw and normalized metrics therefore provide complementary views: raw counts capture the total findings in a solution, whereas normalized metrics capture their concentration relative to code size.

Overall, RQ1 shows that functional performance alone does not determine the static quality profile of generated code. Models exhibit distinct maintainability-related, structural, and reliability-related characteristics, including among functionally correct programs. Moreover, the apparent increase in static-analysis findings on harder tasks is partly explained by greater code volume: several raw issue counts increase with difficulty while their length-normalized densities remain stable or decrease.

\begin{figure*}[!t]
\centering

\subfloat[Convention]{%
    \includegraphics[width=0.25\linewidth]{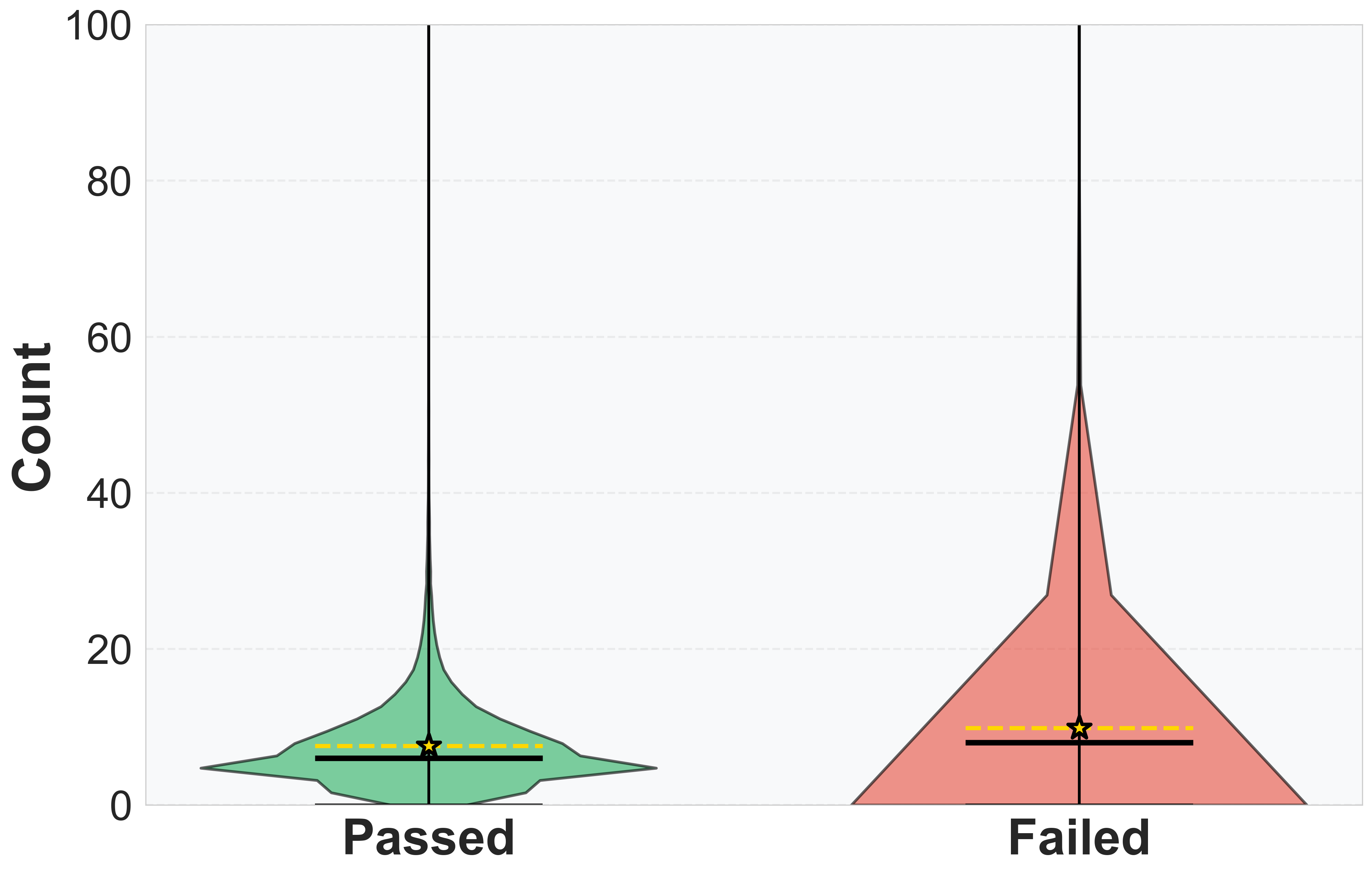}
    \label{fig:box_convention}
}\hspace{0.01\linewidth}%
\includegraphics[height=0.8in]{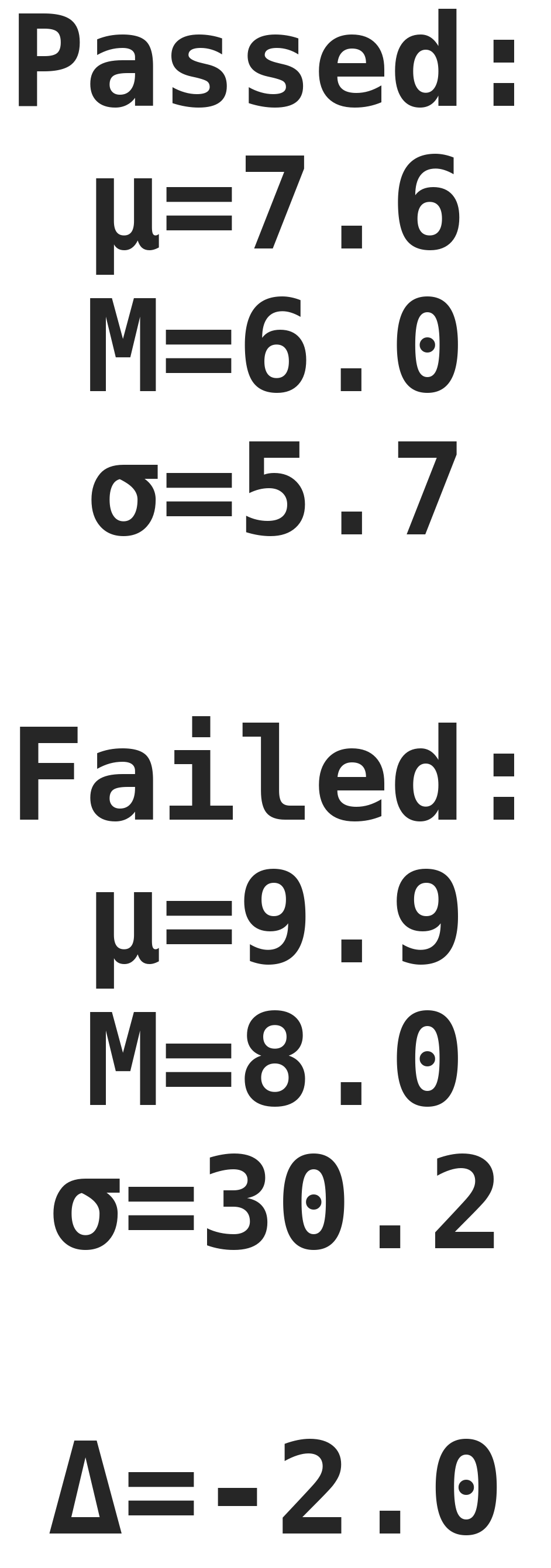}%
\hfill
\subfloat[Warning]{%
    \includegraphics[width=0.25\linewidth]{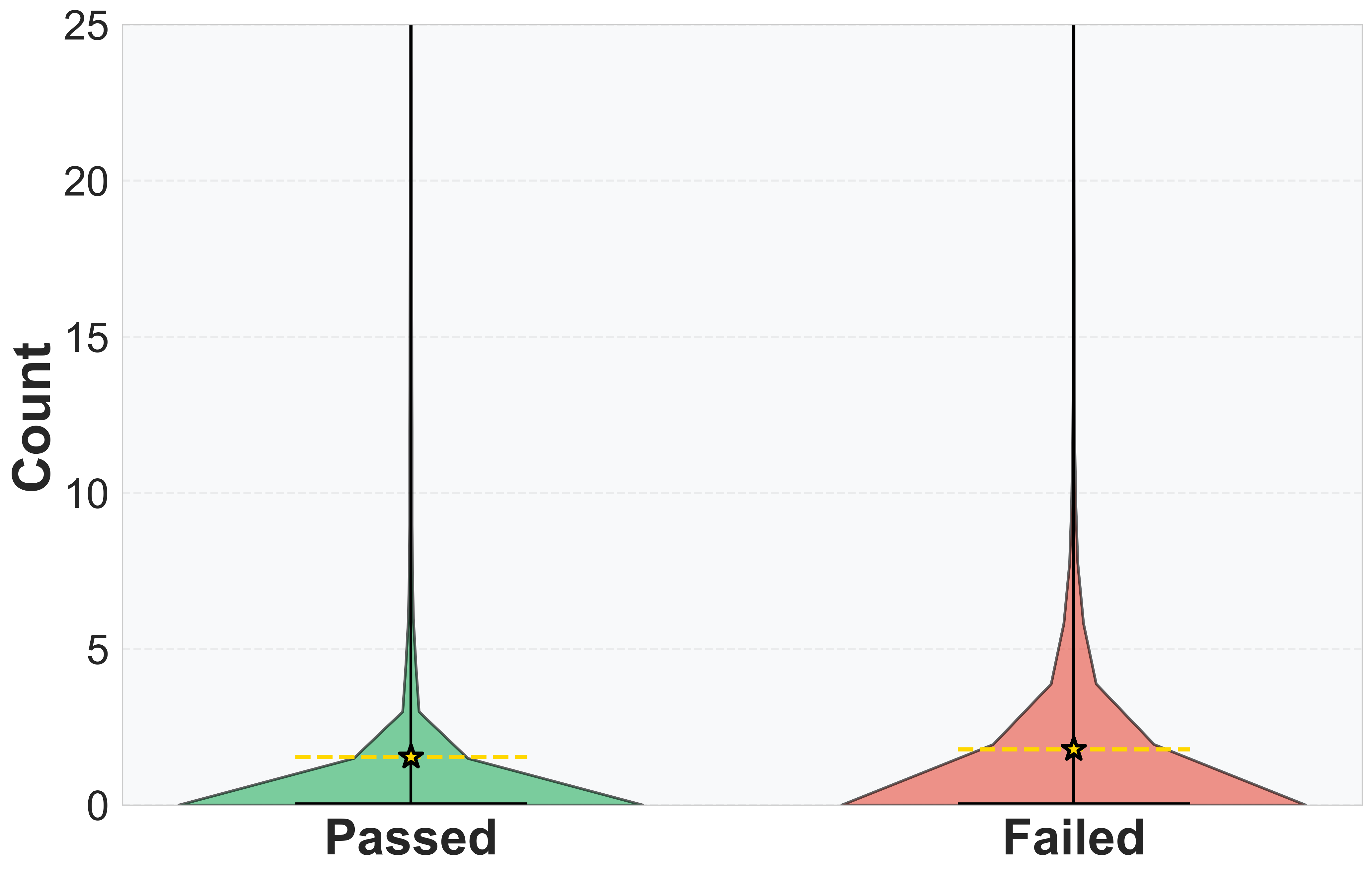}
    \label{fig:box_warning}
}\hspace{0.01\linewidth}%
\includegraphics[height=0.8in]{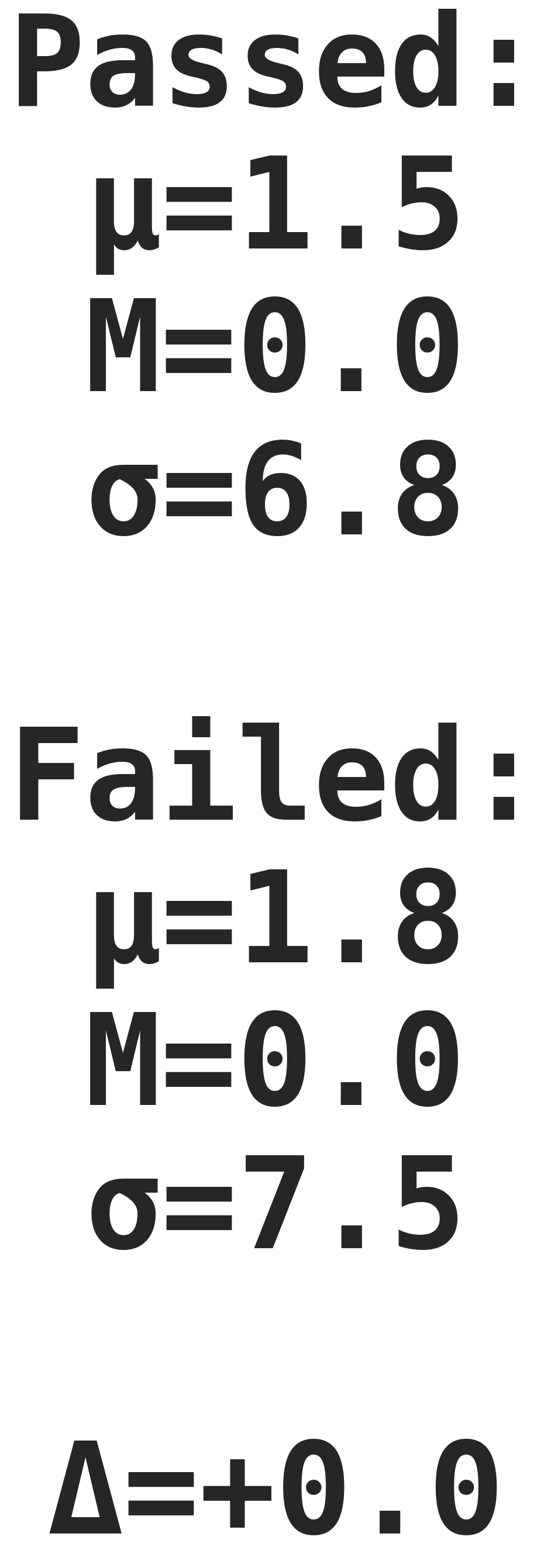}%
\hfill
\subfloat[Refactor]{%
    \includegraphics[width=0.25\linewidth]{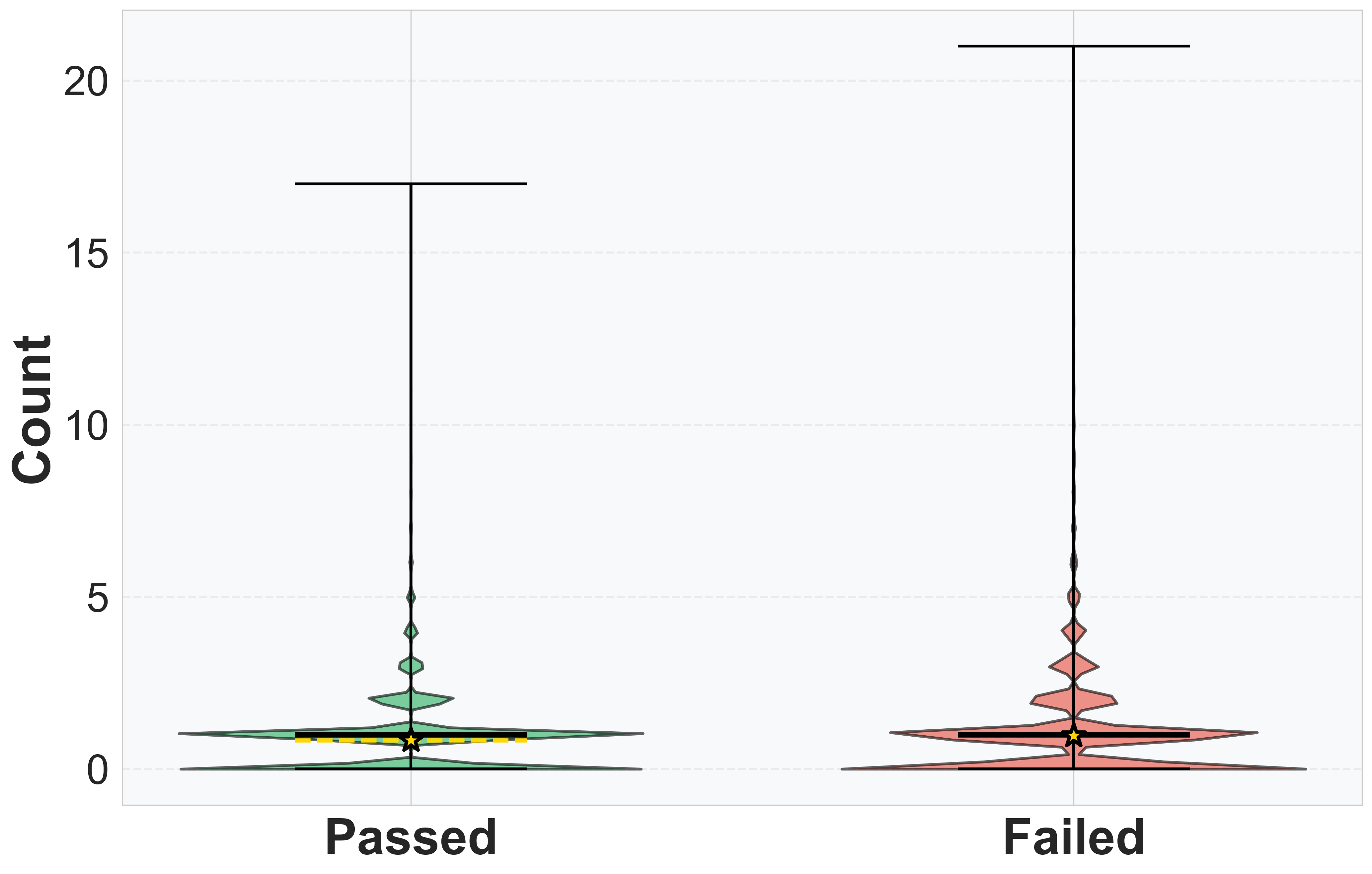}
    \label{fig:box_refactor}
}\hspace{0.01\linewidth}%
\includegraphics[height=0.8in]{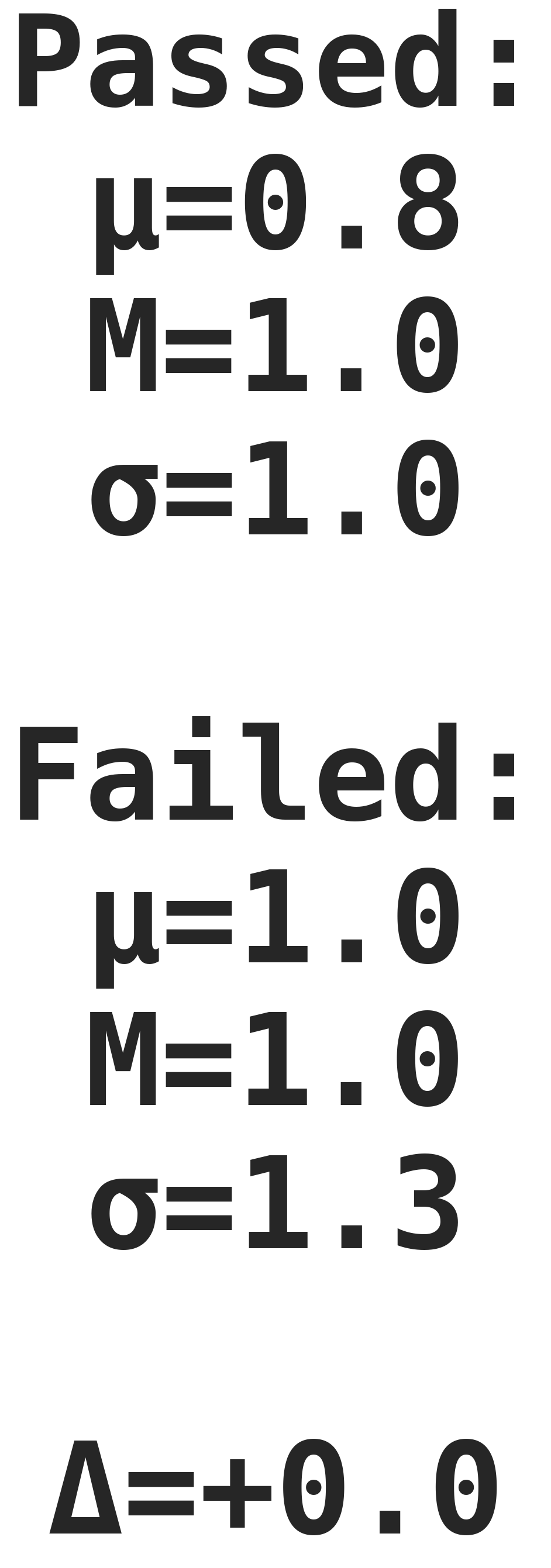}

\vspace{-0.15cm}

\subfloat[Error]{%
    \includegraphics[width=0.25\linewidth]{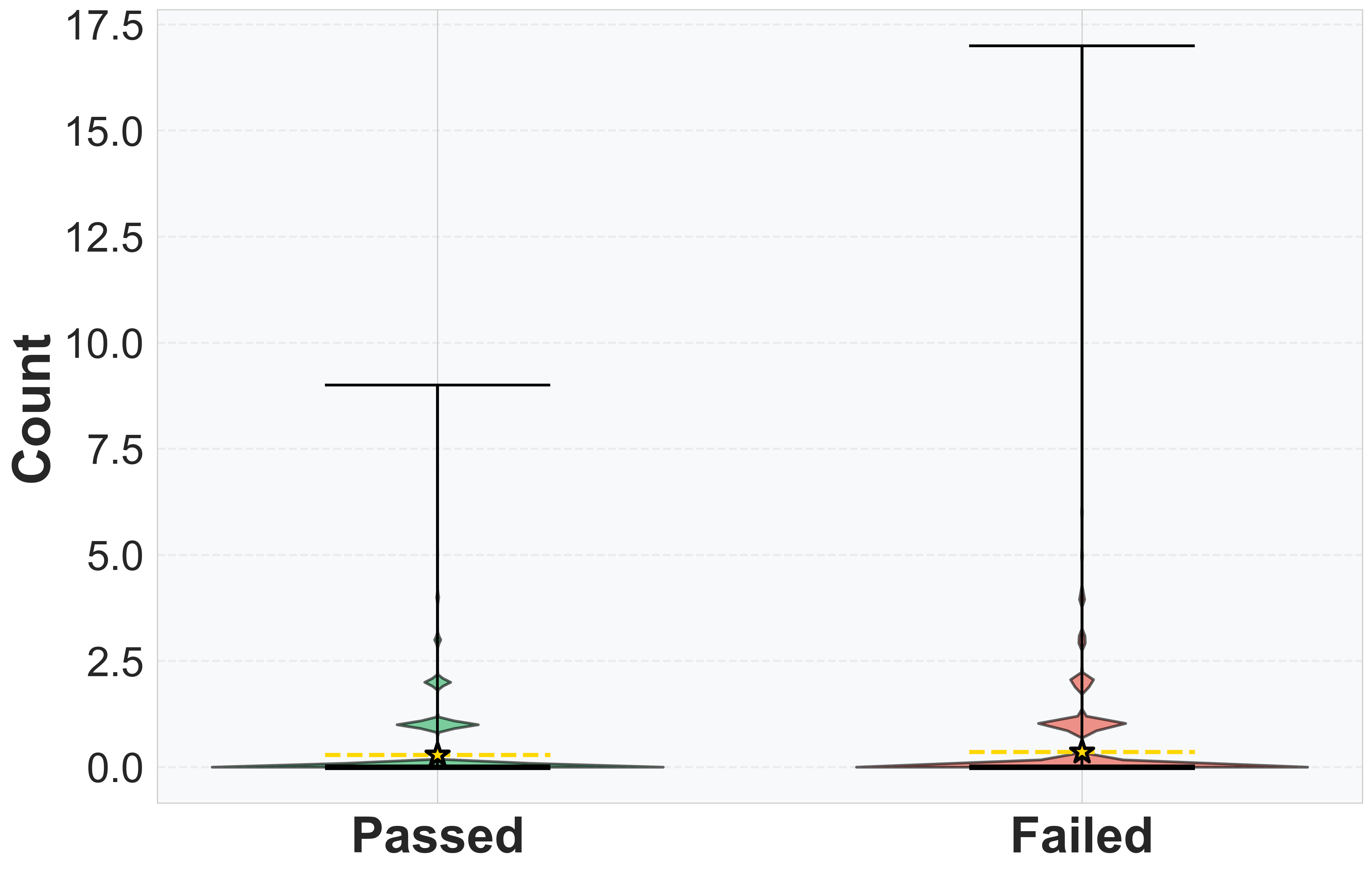}
    \label{fig:box_error}
}\hspace{0.01\linewidth}%
\includegraphics[height=0.8in]{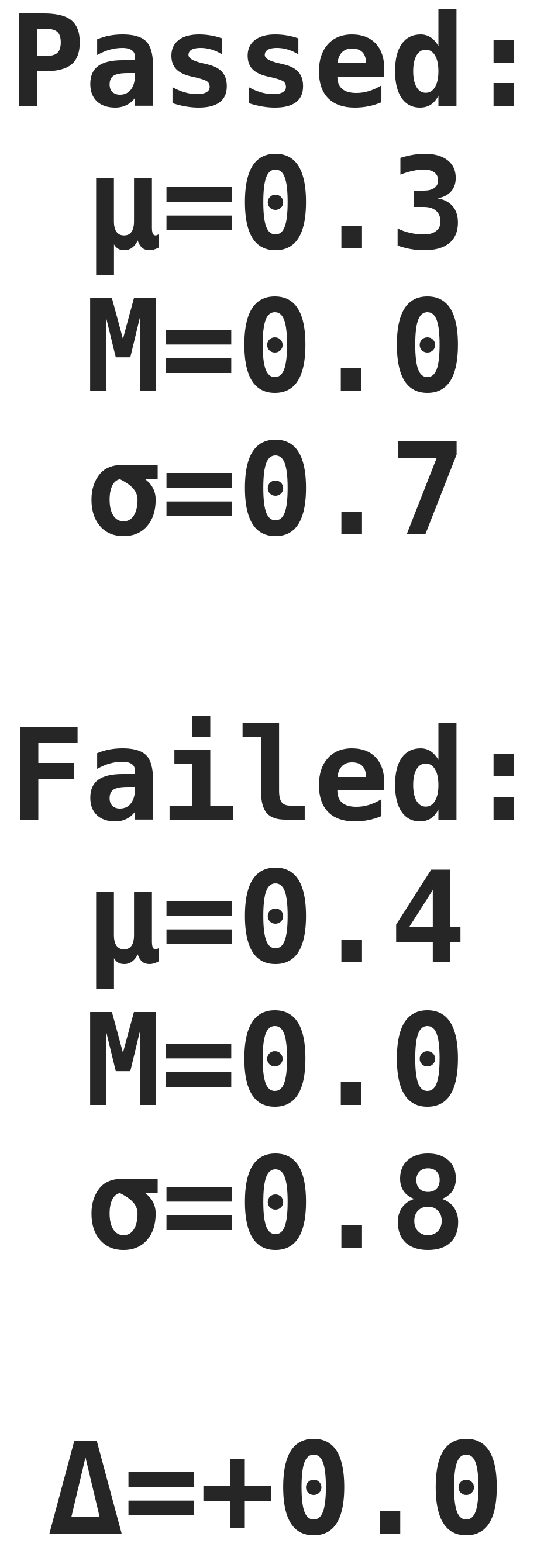}%
\hfill
\subfloat[SLOC]{%
    \includegraphics[width=0.25\linewidth]{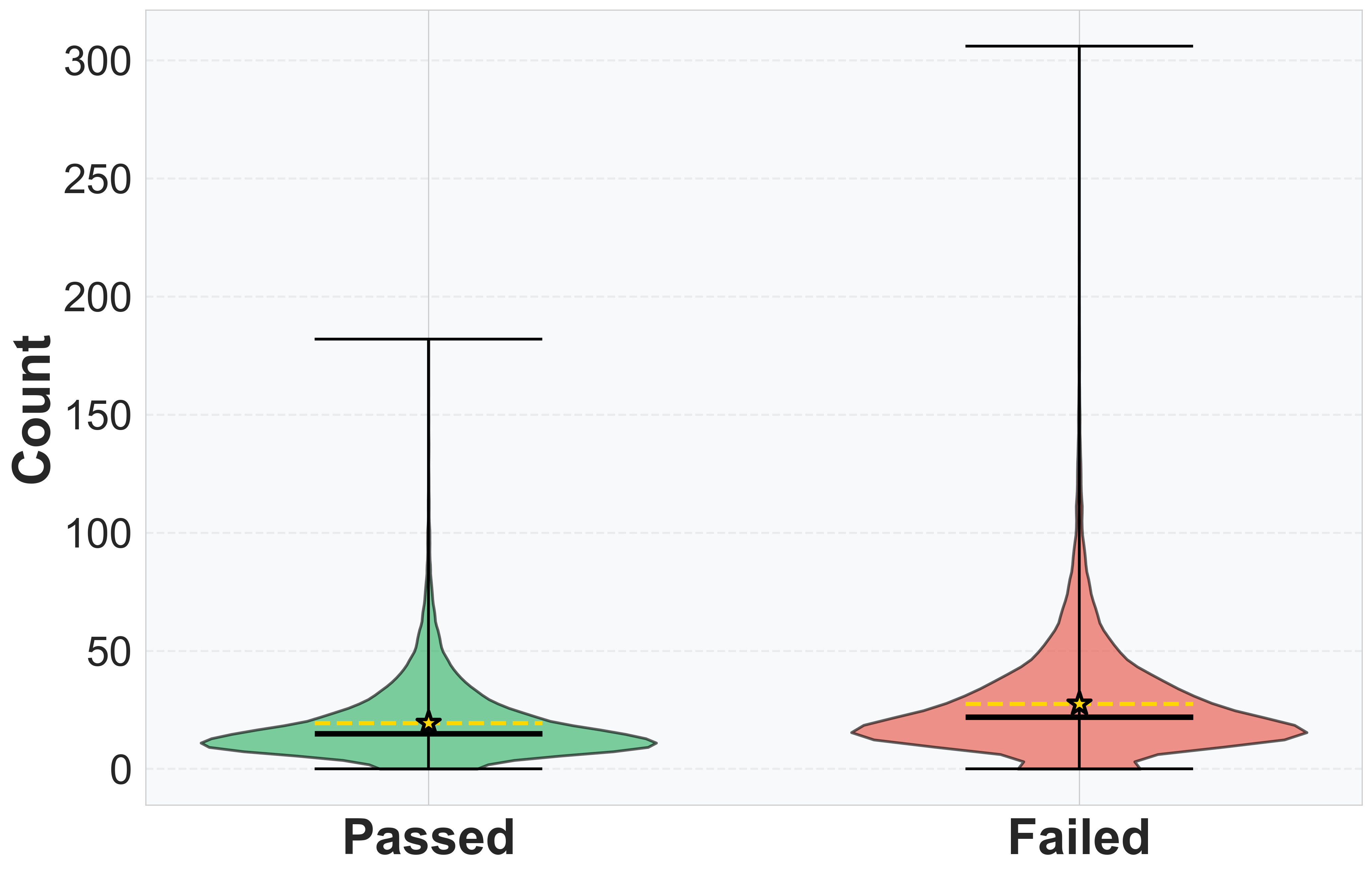}
    \label{fig:box_sloc}
}\hspace{0.01\linewidth}%
\includegraphics[height=0.8in]{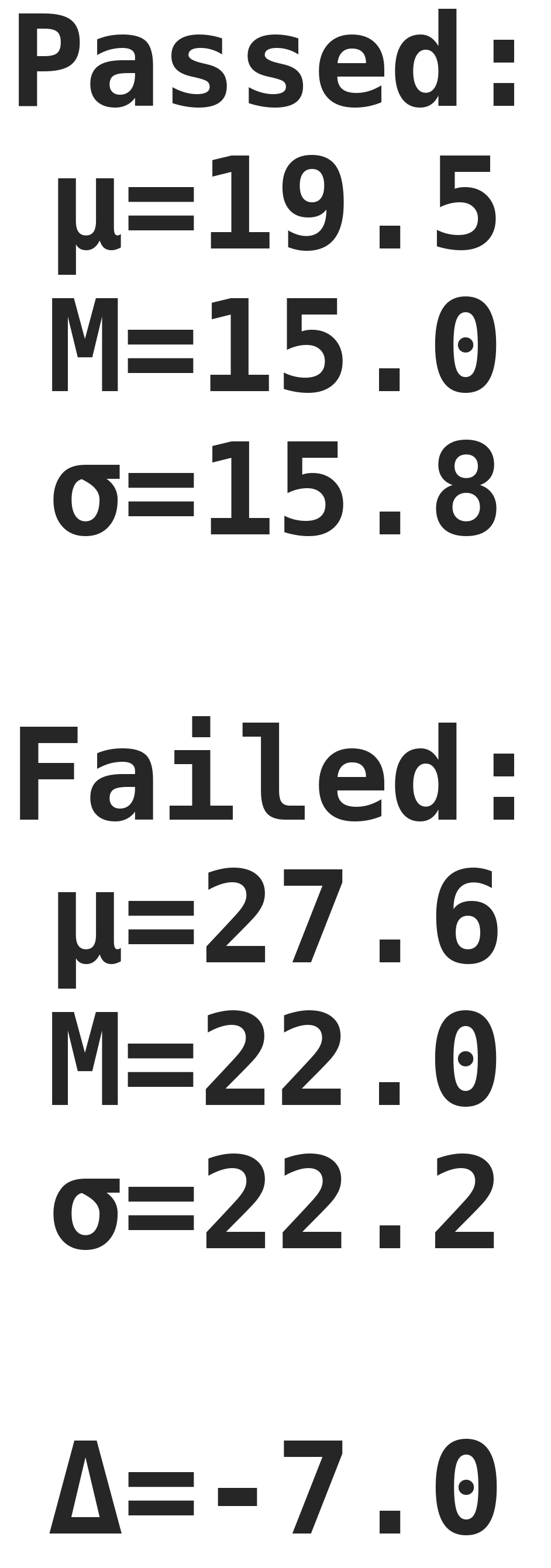}%
\hfill
\subfloat[LLOC]{%
    \includegraphics[width=0.25\linewidth]{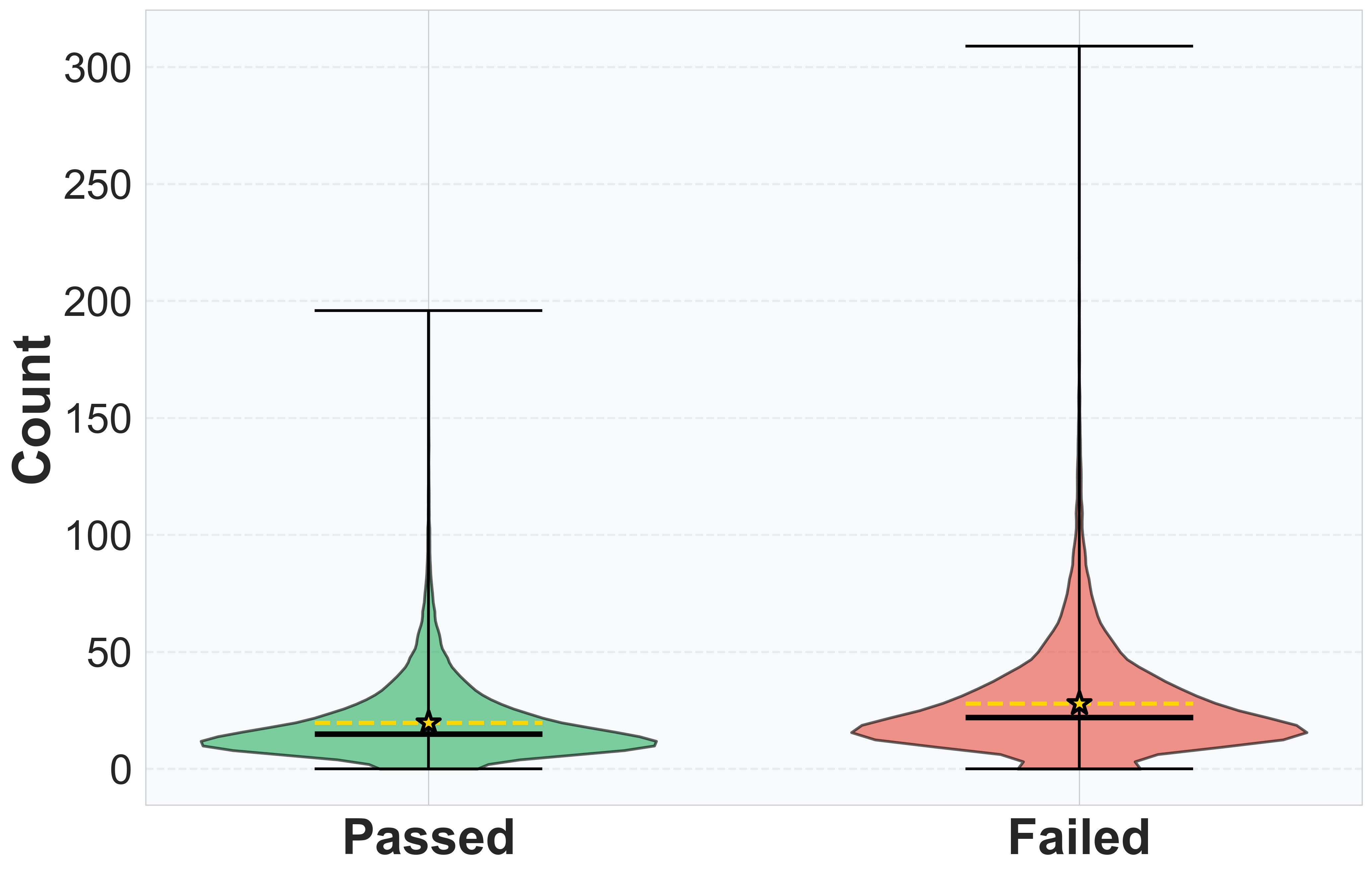}
    \label{fig:box_lloc}
}\hspace{0.01\linewidth}%
\includegraphics[height=0.8in]{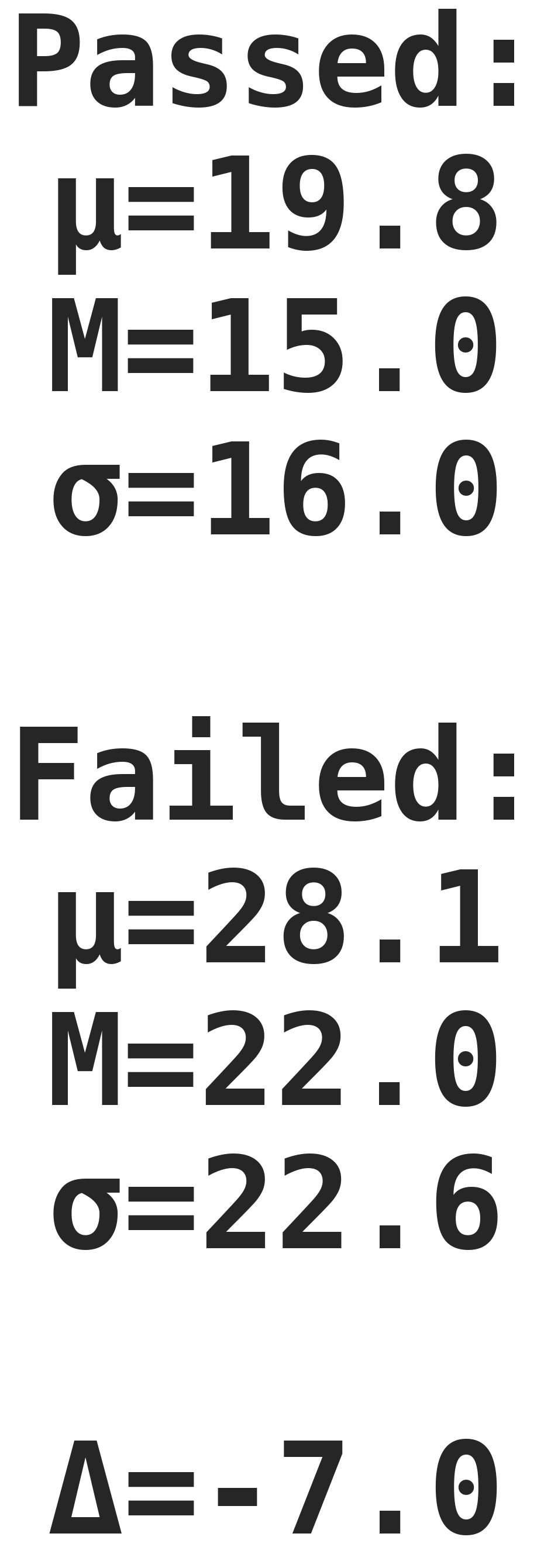}

\vspace{-0.15cm}

\subfloat[Complexity]{%
    \includegraphics[width=0.25\linewidth]{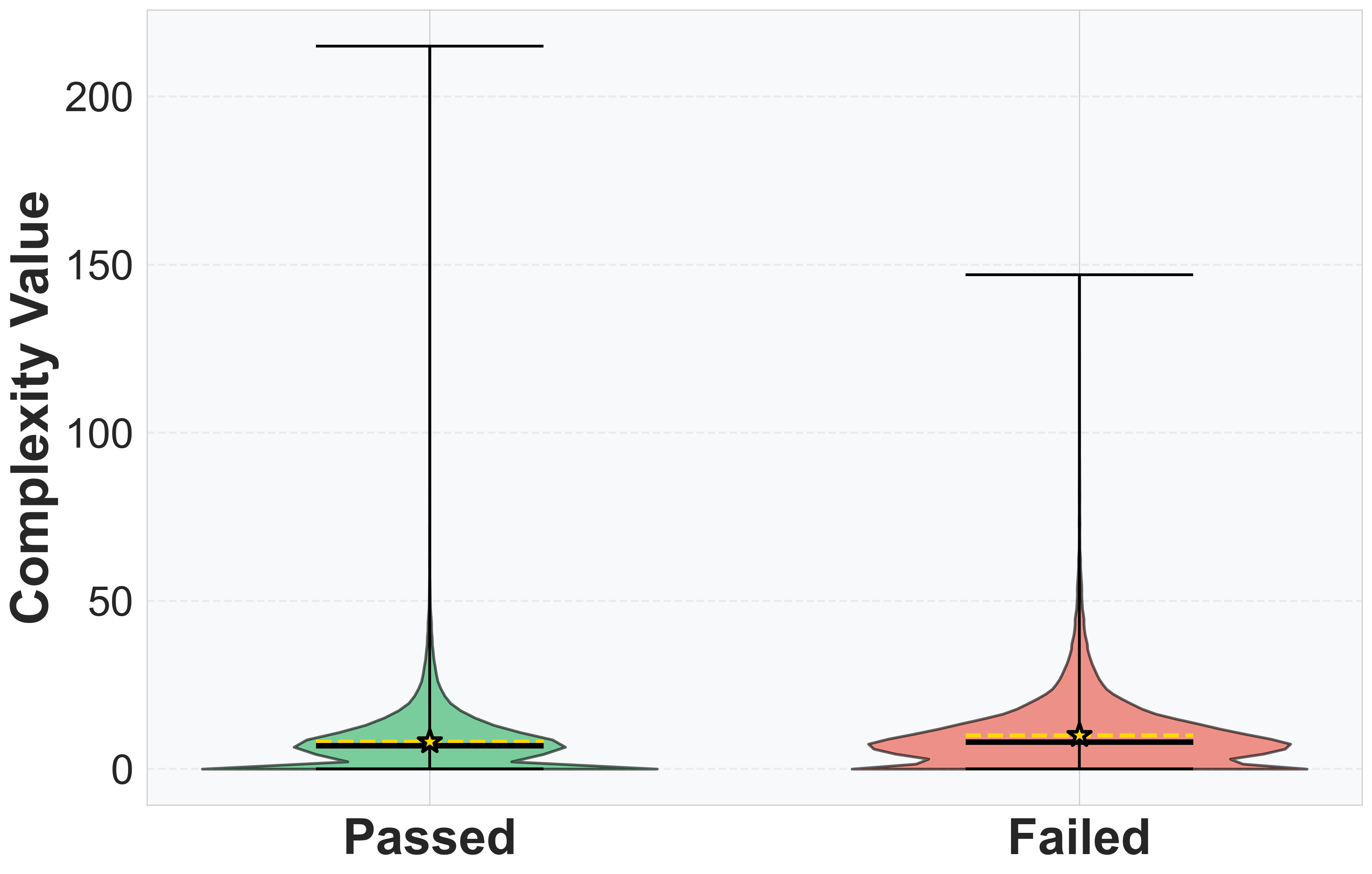}
    \label{fig:box_complexity}
}\hspace{0.01\linewidth}%
\includegraphics[height=0.8in]{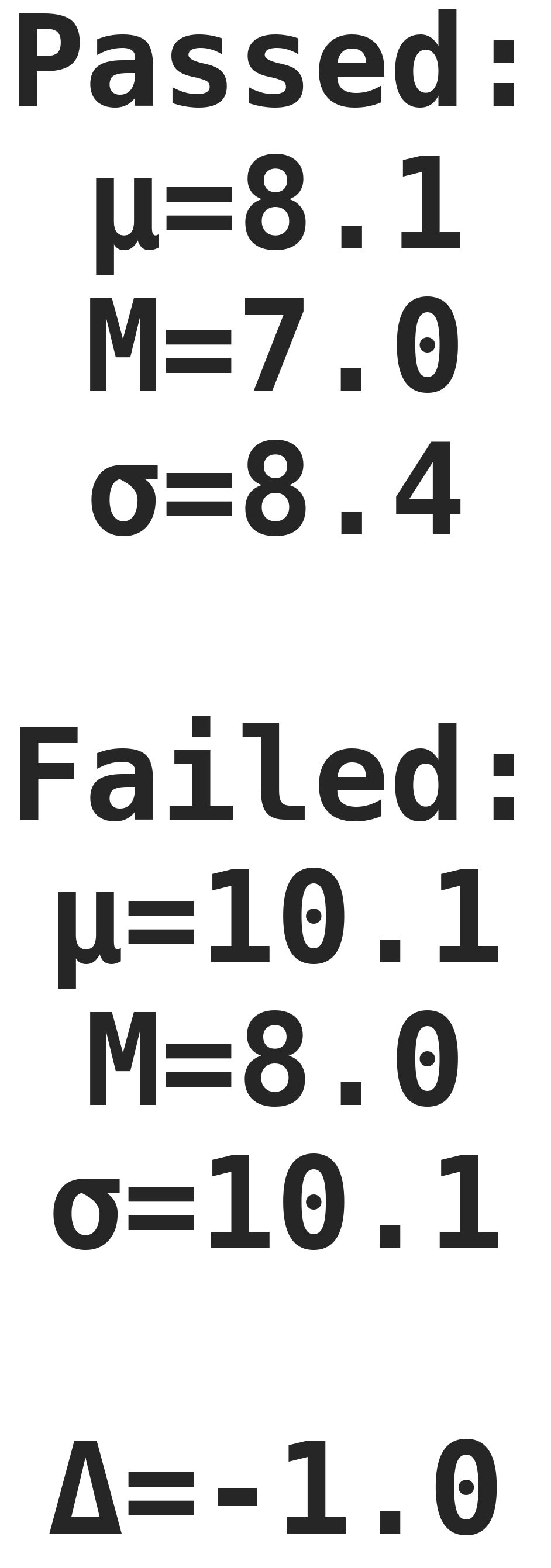}%
\hfill
\subfloat[LOC]{%
    \includegraphics[width=0.25\linewidth]{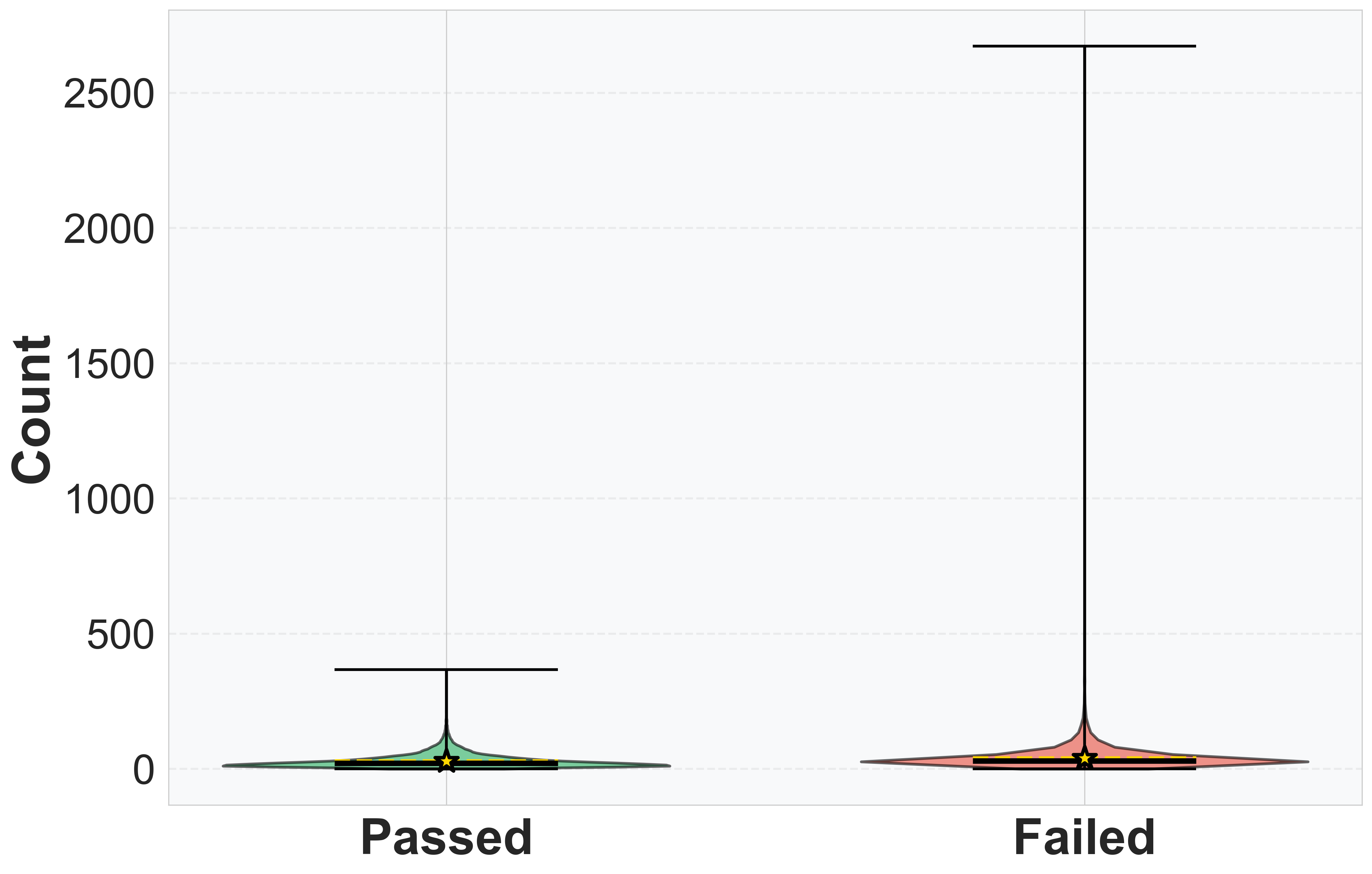}
    \label{fig:box_loc}
}\hspace{0.01\linewidth}%
\includegraphics[height=0.8in]{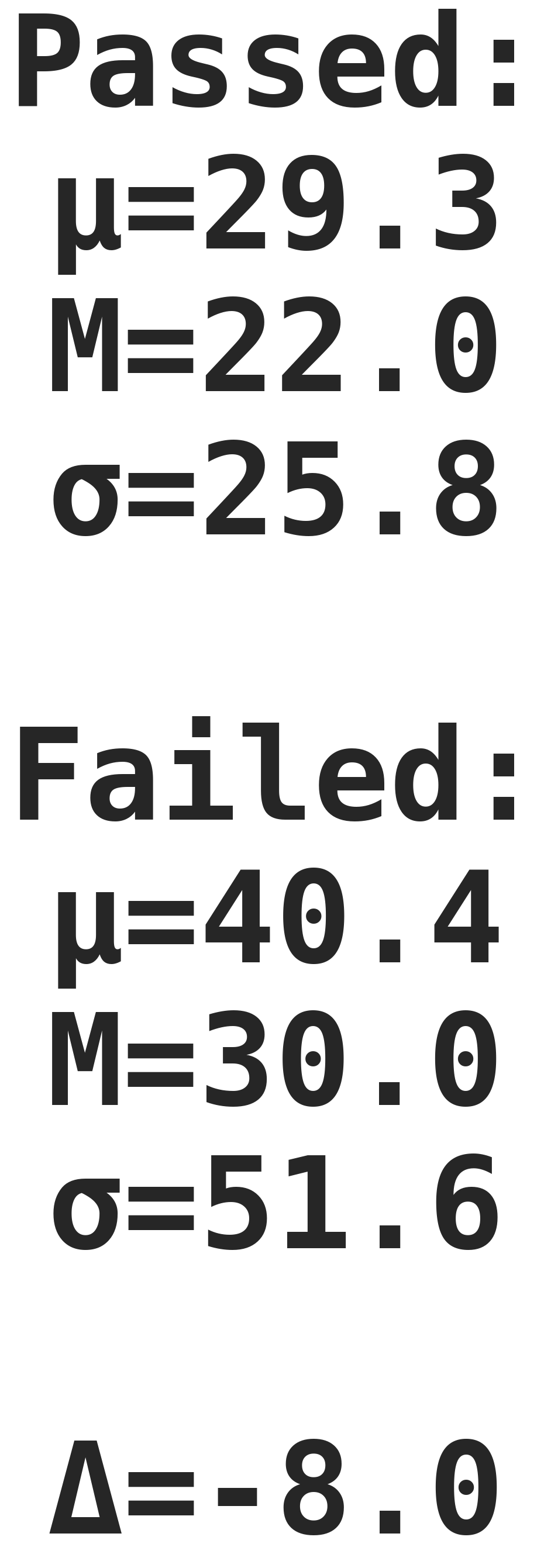}%
\hfill
\subfloat[Comments]{%
    \includegraphics[width=0.25\linewidth]{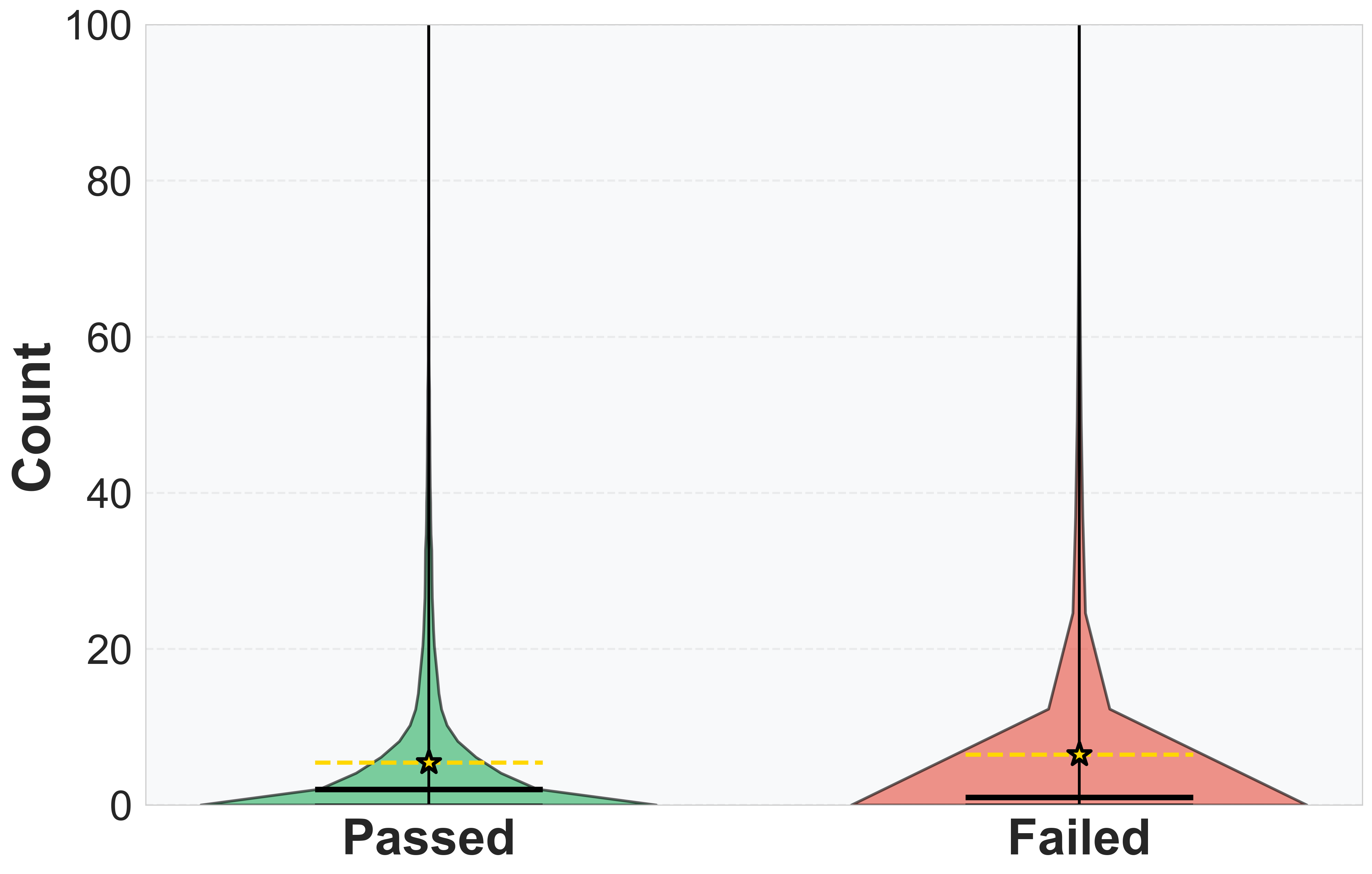}
    \label{fig:box_comments}
}\hspace{0.01\linewidth}%
\includegraphics[height=0.8in]{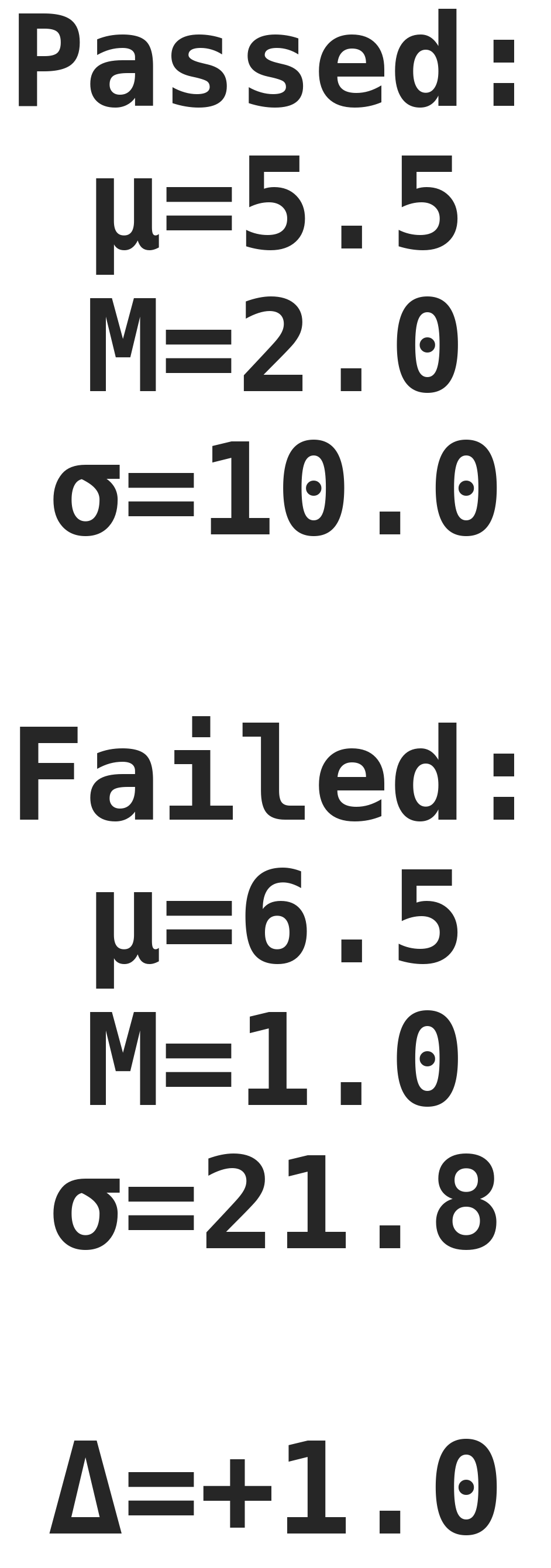}

\vspace{-0.1cm}

\subfloat[Blank Lines]{%
    \includegraphics[width=0.25\linewidth]{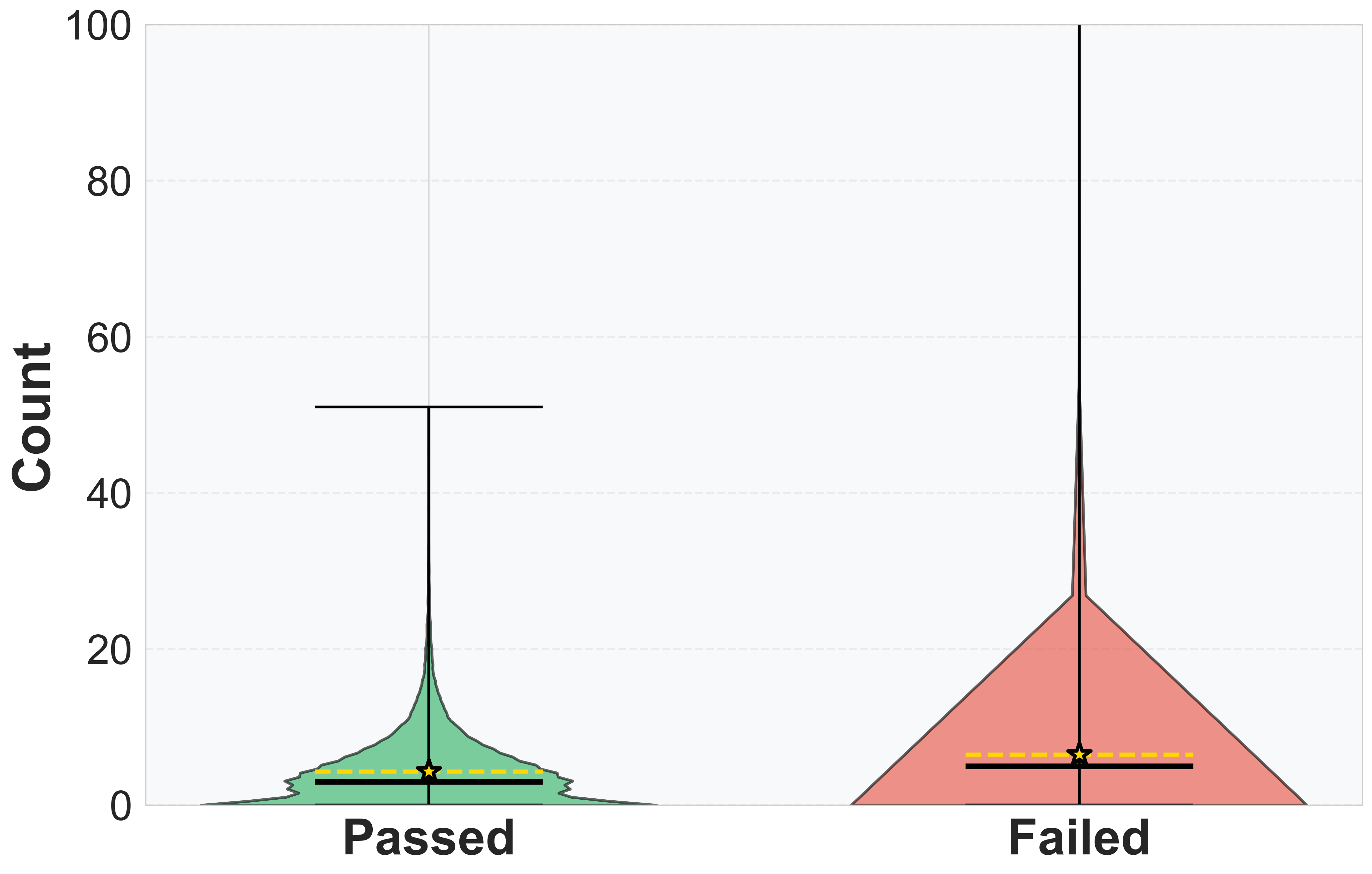}
    \label{fig:box_blank}
}\hspace{0.01\linewidth}%
\includegraphics[height=0.8in]{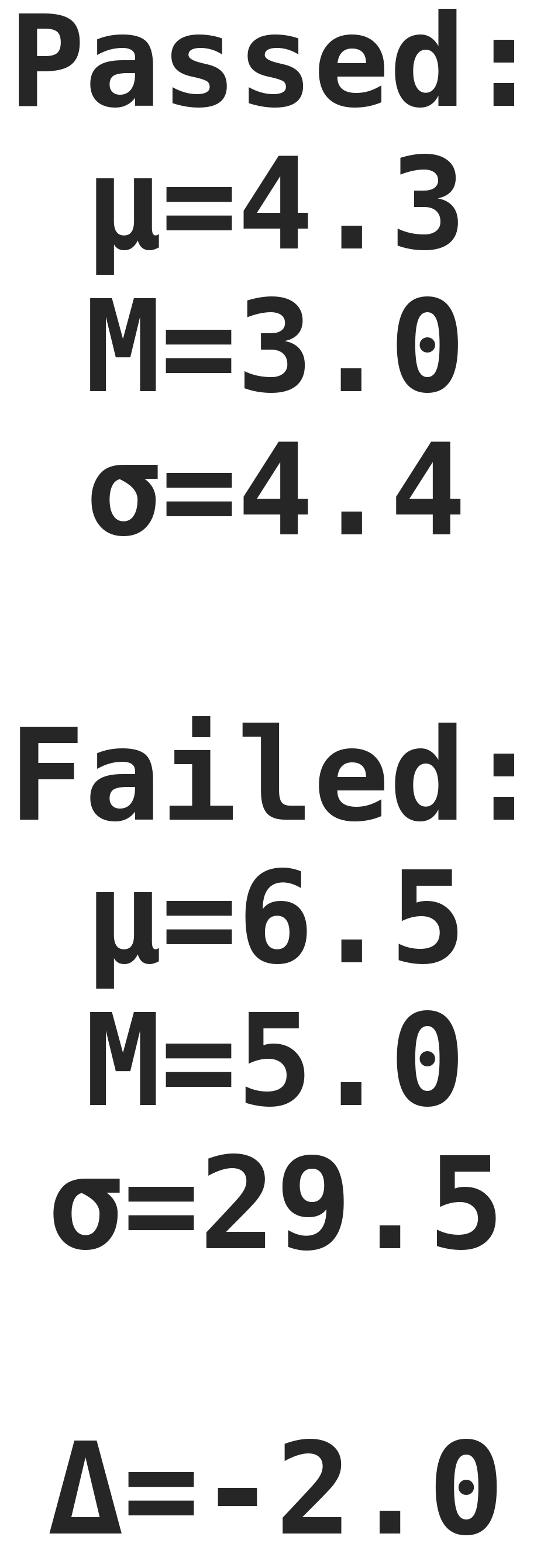}%
\hfill
\subfloat[Functions]{%
    \includegraphics[width=0.25\linewidth]{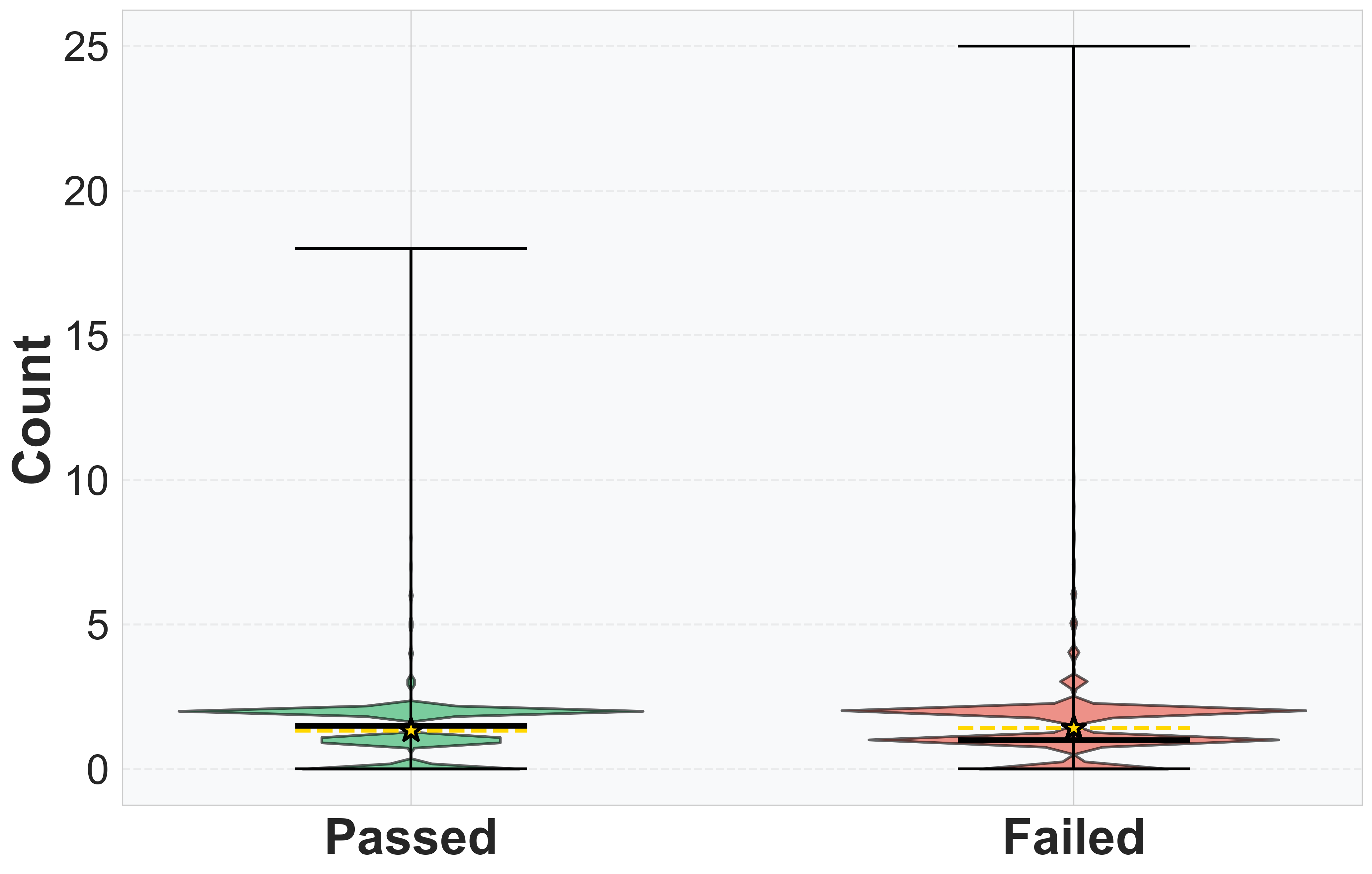}
    \label{fig:box_functions}
}\hspace{0.01\linewidth}%
\includegraphics[height=0.8in]{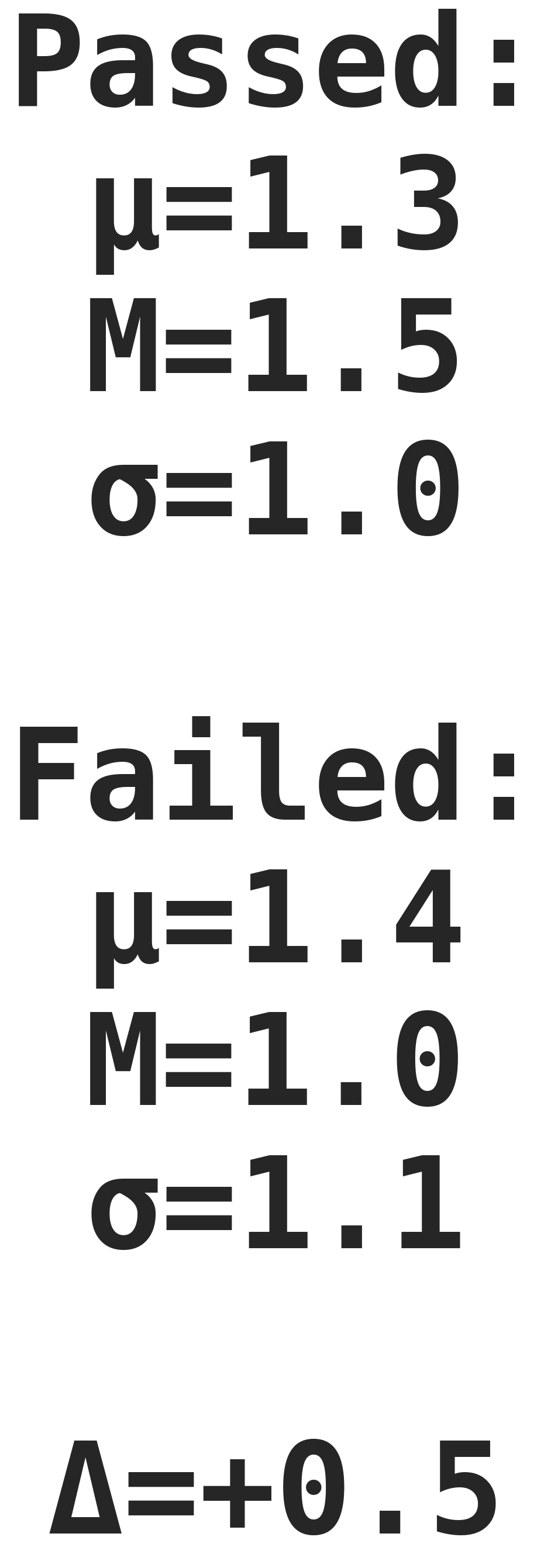}%
\hfill
\subfloat[Maintainability Index]{%
    \includegraphics[width=0.25\linewidth]{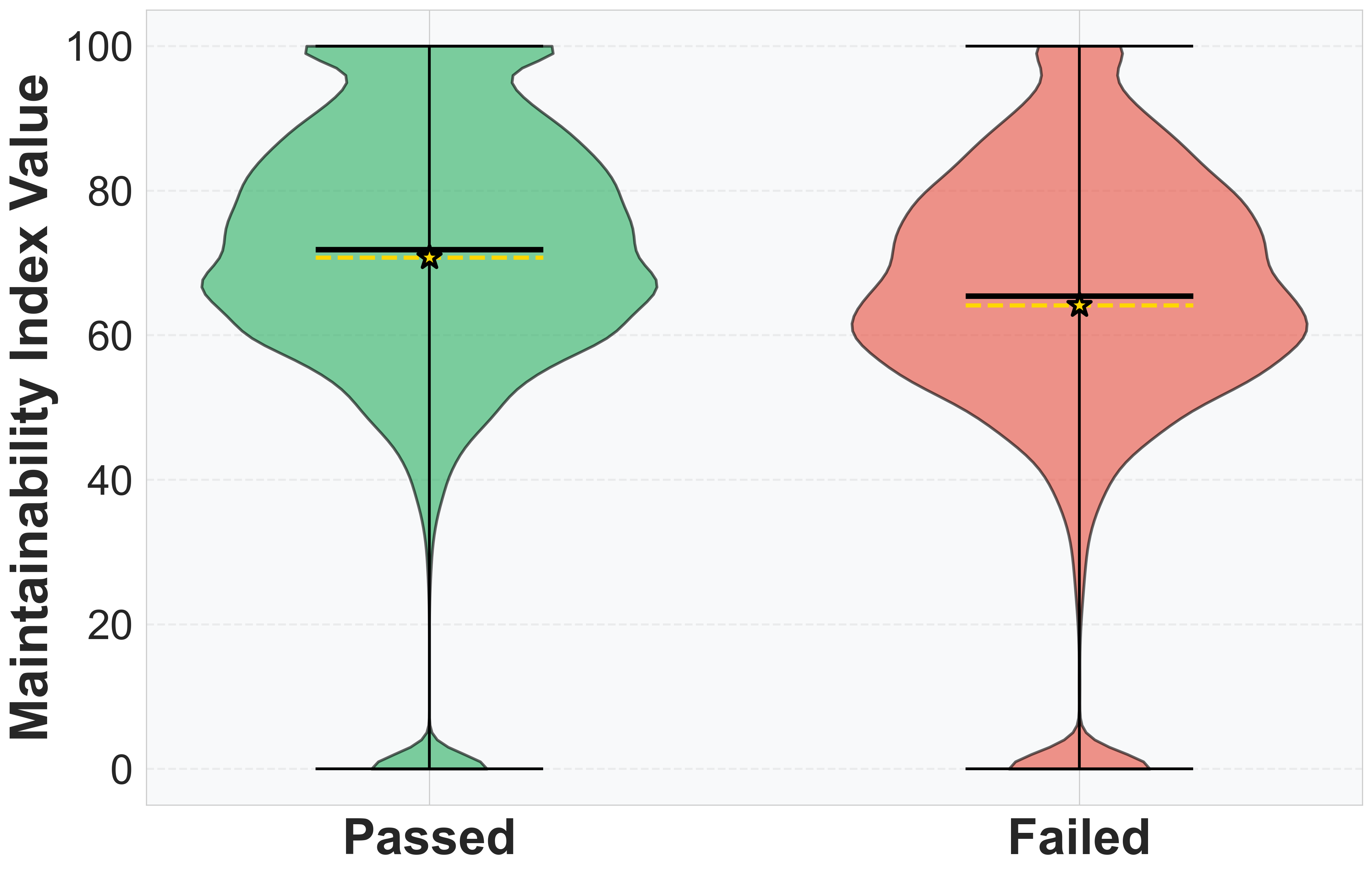}
    \label{fig:box_maint_index}
}\hspace{0.01\linewidth}%
\includegraphics[height=0.8in]{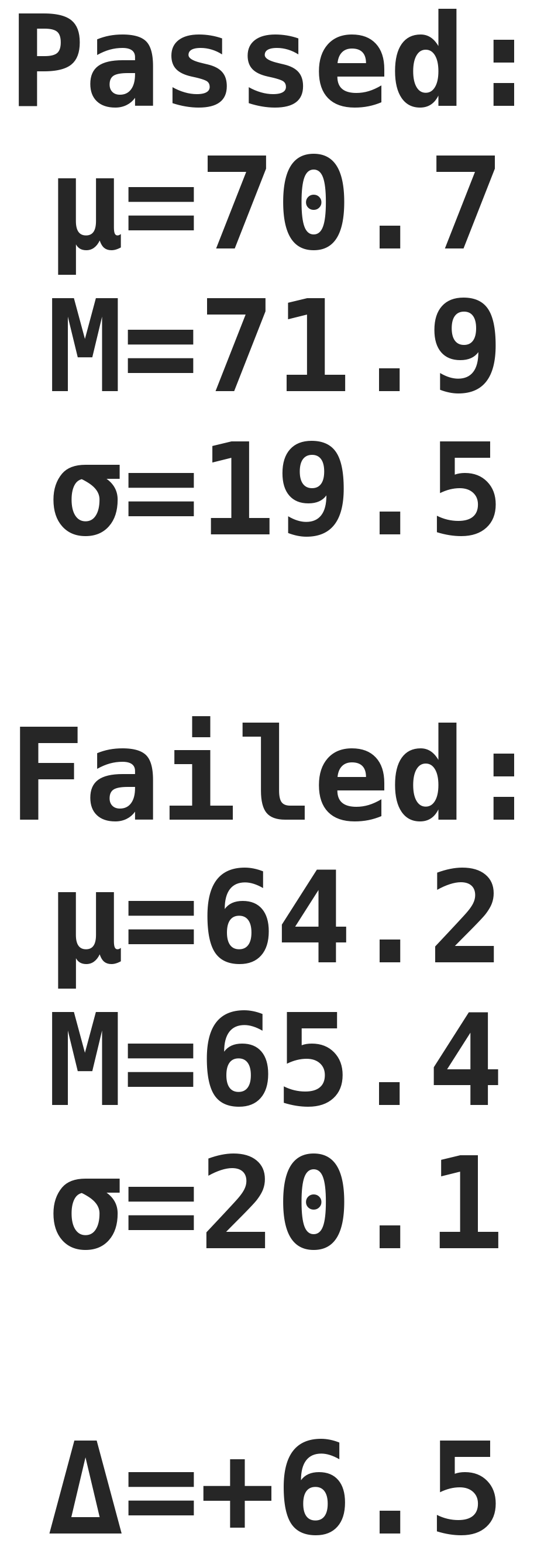}

\vspace{-0.15cm}

\subfloat[Maintainability Issues]{%
    \includegraphics[width=0.25\linewidth]{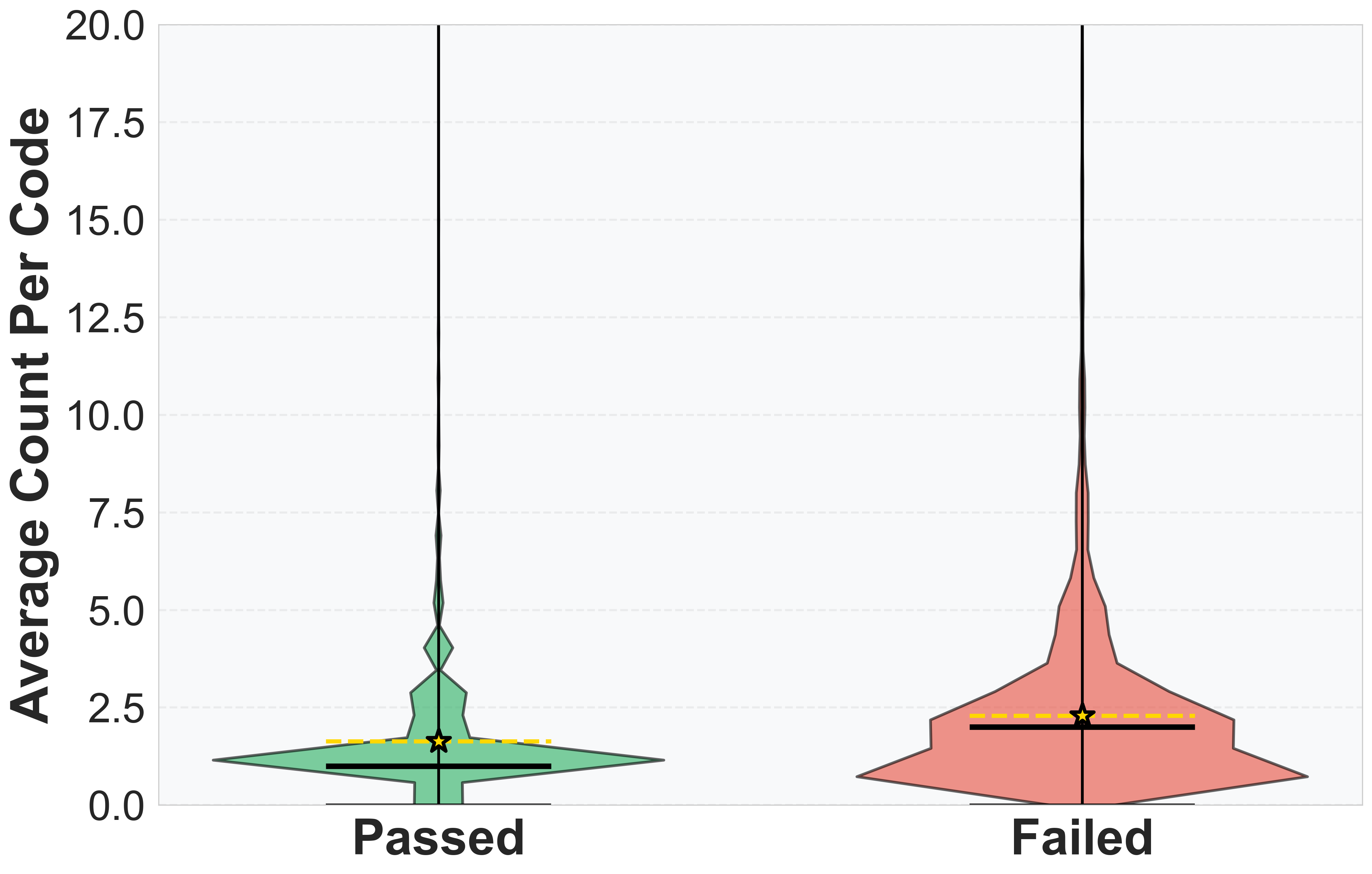}
    \label{fig:box_maint_issues}
}\hspace{0.01\linewidth}%
\includegraphics[height=0.8in]{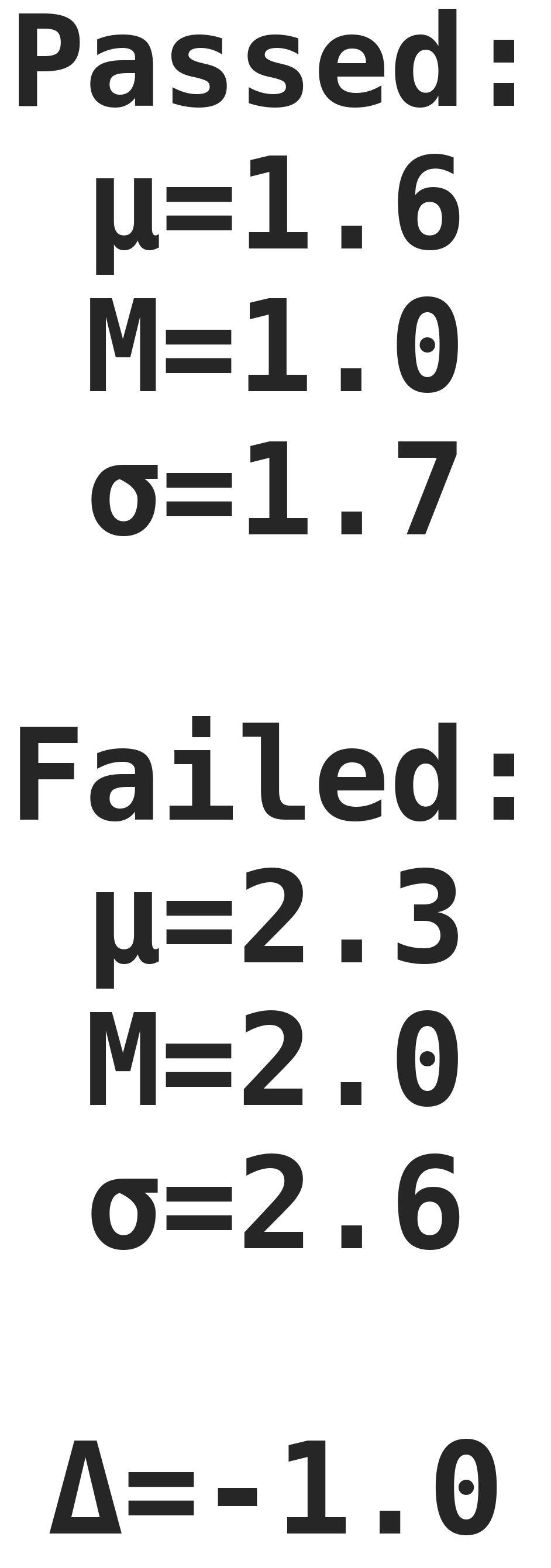}%
\hfill
\subfloat[Reliability Issues]{%
    \includegraphics[width=0.25\linewidth]{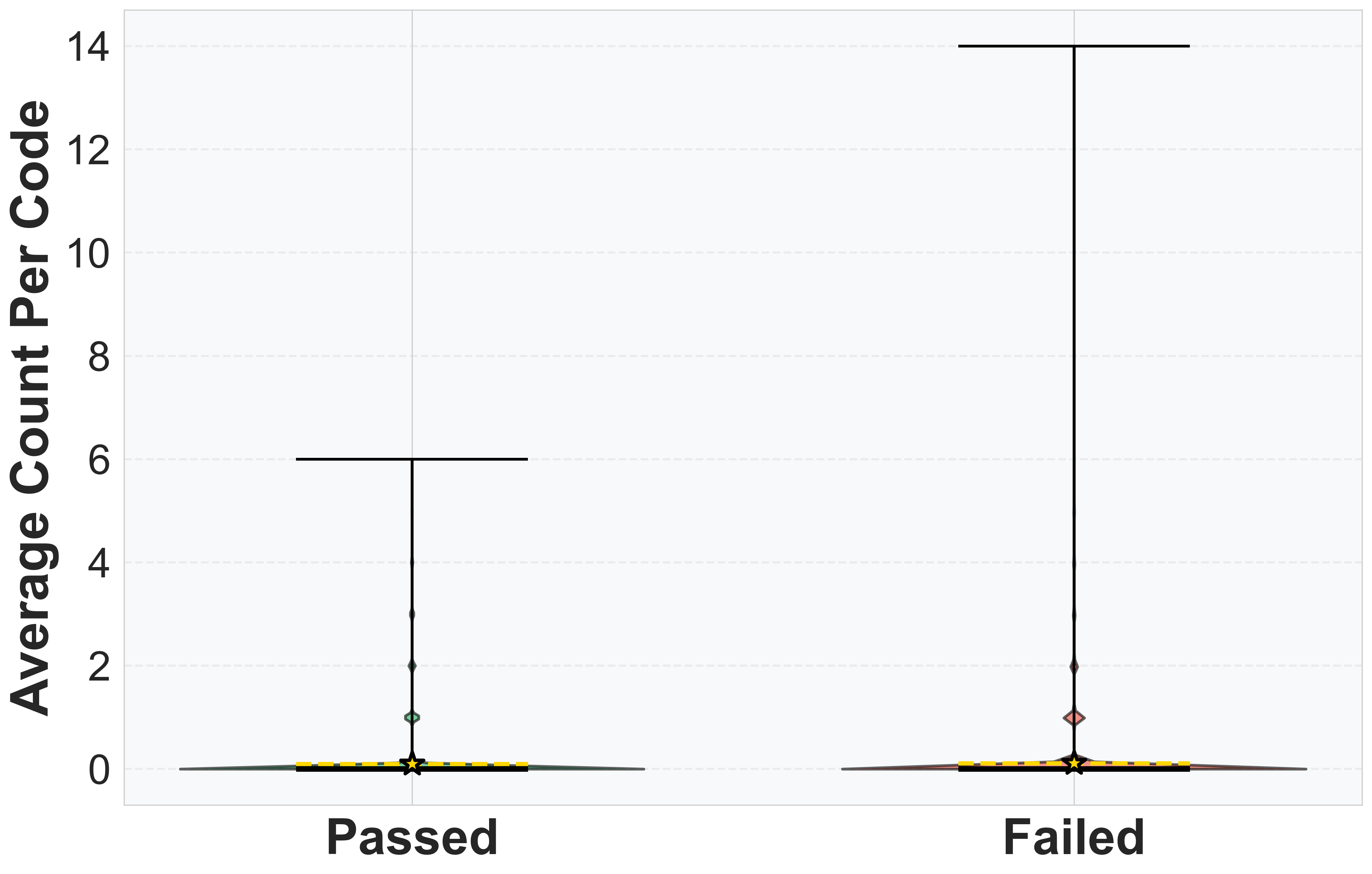}
    \label{fig:box_rel_issues}
}\hspace{0.01\linewidth}%
\includegraphics[height=0.8in]{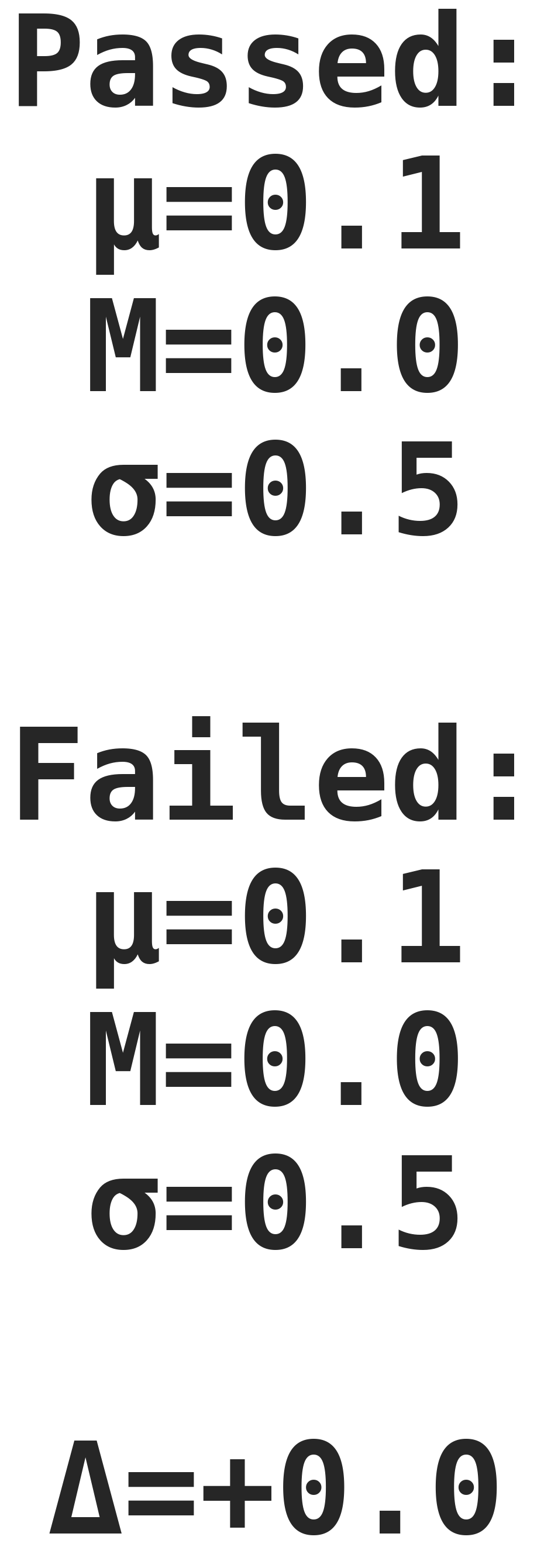}%
\hfill
\subfloat[Security Issues]{%
    \includegraphics[width=0.25\linewidth]{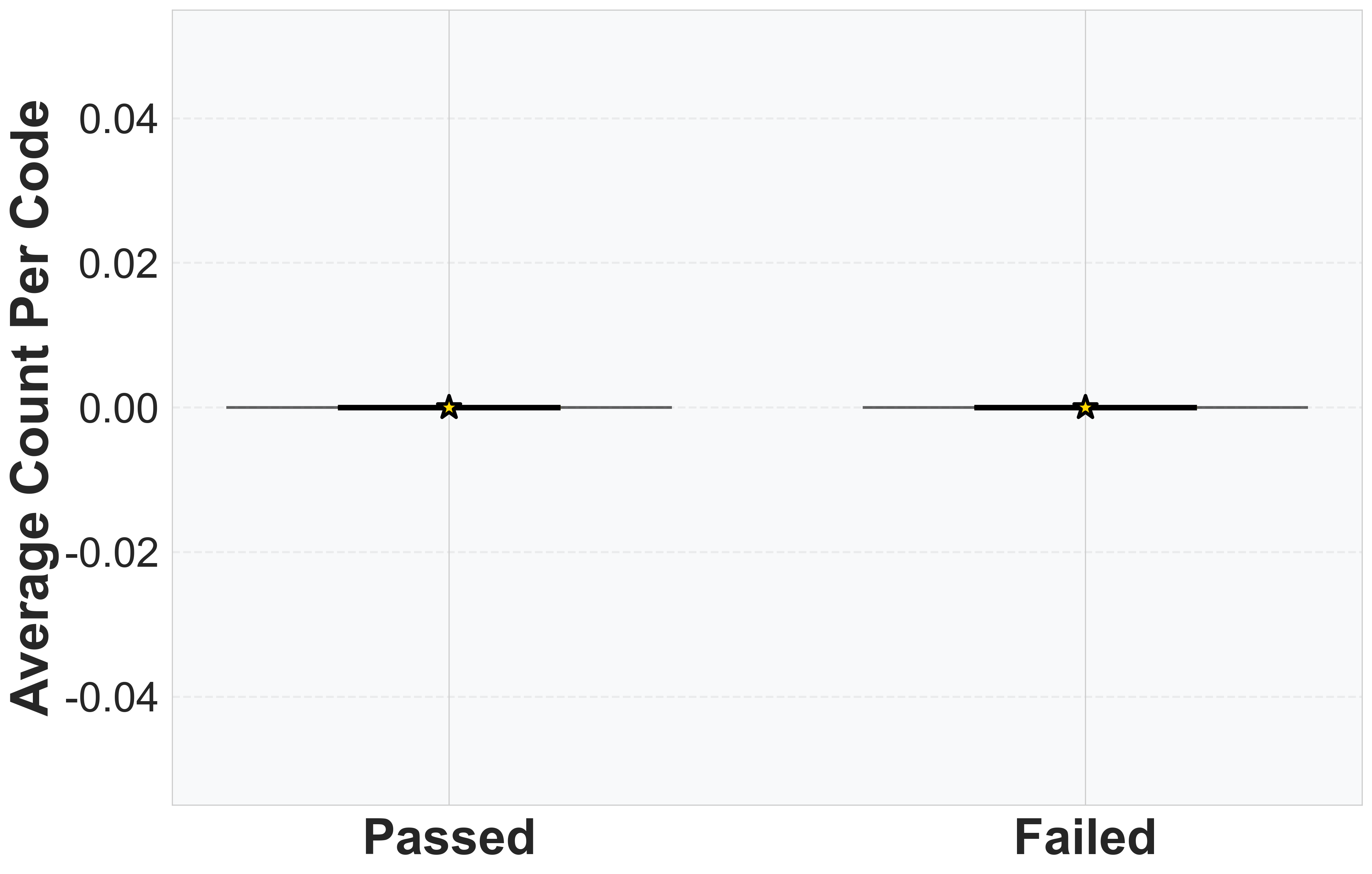}
    \label{fig:box_sec_issues}
}\hspace{0.01\linewidth}%
\includegraphics[height=0.8in]{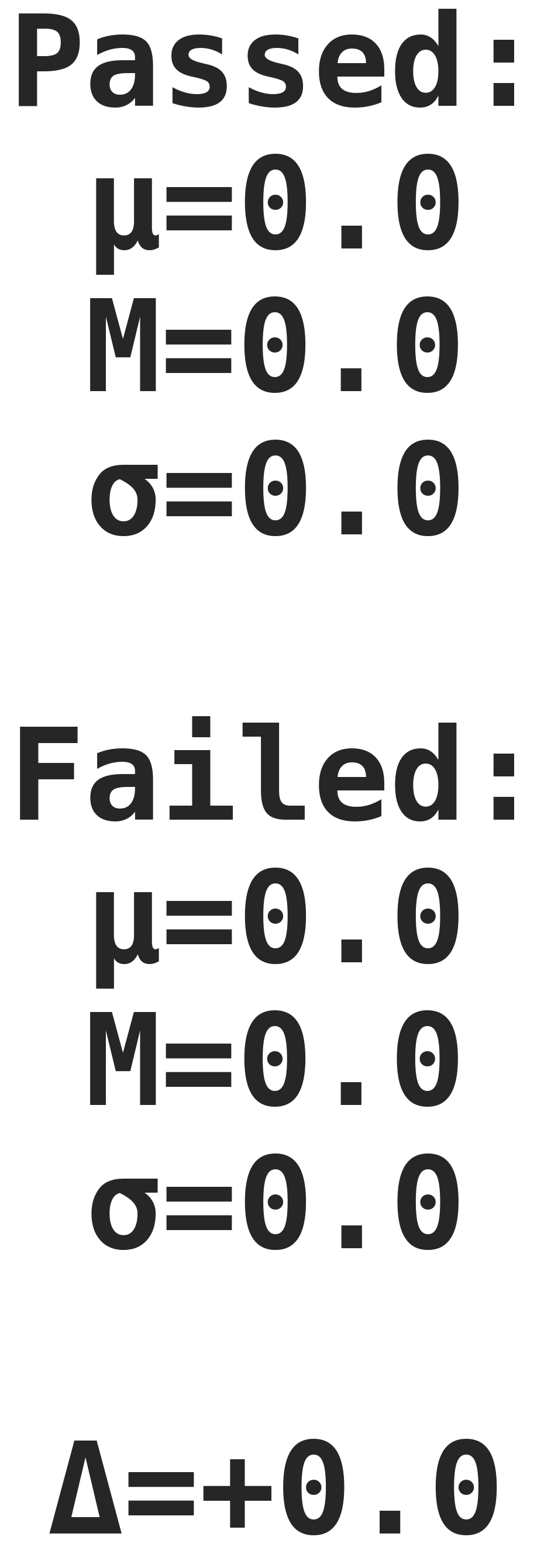}

\vspace{0.1cm}

\raisebox{0pt}[\dimexpr\height-0.5\baselineskip][\depth]{%
    \includegraphics[width=0.60\linewidth]{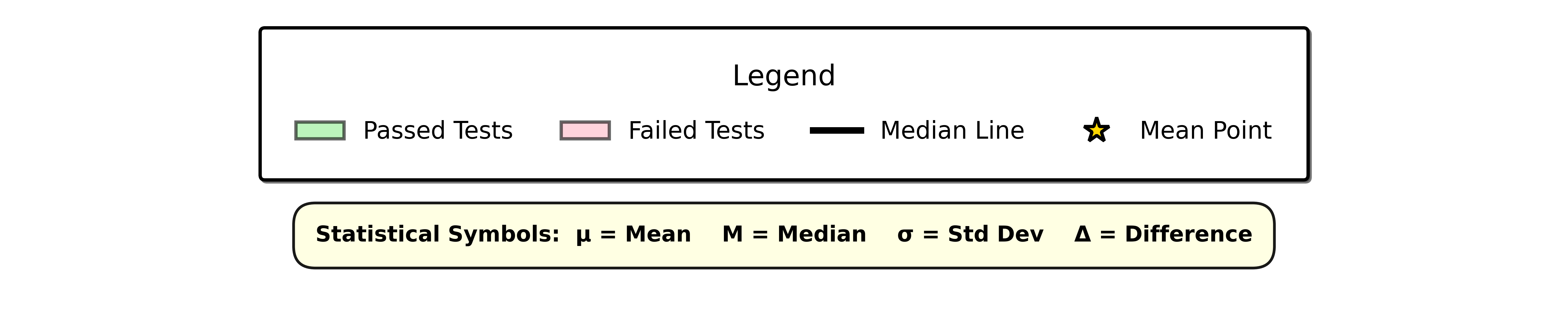}%
}

\caption{
Distribution of raw static code quality metrics comparing code samples that pass the LiveCodeBench correctness tests (green) with those that fail (red).
}
\label{fig:boxplots_all_metrics}
\end{figure*}

\subsection{RQ2: What trade-offs emerge between functional correctness and static code quality metrics in LLM-generated code?}

To answer RQ2, we examine how functional correctness relates to the multi-dimensional static quality profiles identified in RQ1, at both the model level and the sample level, as illustrated in Figure~\ref{fig:boxplots_all_metrics}. Two broad patterns emerge from the raw metrics. First, correctness is often associated with \emph{larger and more explicitly structured code}. At the model level, higher-correctness models generally show higher values of \texttt{SLOC}, \texttt{LOC}, and \texttt{LLOC}, and often also higher \texttt{Comments} and \texttt{Blank Lines}, suggesting that stronger functional performance is frequently linked to more complete and explicitly organized implementations. A similar pattern appears at the sample level: code that passes the benchmark tests tends to be longer on average and also shows somewhat higher \texttt{Comments} and \texttt{Blank Lines} than code that fails. This difference is consistent with the possibility that some failed solutions contain incomplete or partially developed implementations, rather than indicating that shorter code is inherently preferable.

Second, correctness does not simply correspond to minimal or uniformly simpler code. Higher-correctness models often also exhibit higher \texttt{Average Comple\-xity}, indicating that better functional performance is commonly associated with more fully developed control flow and implementation logic rather than shorter or simpler solutions. At the same time, passed code shows a maintainability-related advantage at the sample level: it has a higher \texttt{Maintainability Index} and fewer SonarQube \texttt{Maintainability Issues}. This indicates that correct solutions receive more favorable maintainability-related static-analysis scores on average, even though achieving correctness may require larger and more structurally involved implementations.

Among the Pylint metrics, \texttt{Convention} and \texttt{Refactor} tend to be higher for stronger models, which is consistent with the fact that longer and more complex implementations create more opportunities for style- and refactor-related findings. By contrast, \texttt{Error} generally decreases as correctness increases, although the pattern is not strictly monotonic, and \texttt{Warning} remains less stable across models. At the sample level, \texttt{Convention} provides the clearest separation between passed and failed code, whereas \texttt{Warning}, \texttt{Refactor}, and \texttt{Error} overlap more strongly. The \texttt{Functions} metric differs only slightly between passed and failed code, indicating that function count is not a major distinguishing factor in this dataset.

The supporting correct-only and length-normalized results qualify these raw relationships. Cross-model differences in size, complexity, and maintainability remain visible among correct programs, showing that they are not entirely caused by incomplete failed outputs. At the same time, several differences in raw issue counts become smaller or change direction after normalization, indicating that part of the apparent correctness--quality trade-off reflects implementation size rather than higher issue concentration.

Overall, RQ2 indicates that functional correctness and static code quality are related but not interchangeable dimensions of generated-code behavior. Correct solutions are often more complete, longer, and more structurally developed, yet higher correctness does not consistently correspond to lower complexity, fewer static-analysis findings, or better maintainability. Functional correctness and static quality should therefore be evaluated as complementary rather than substitutable dimensions.

\section{Discussion}

Three findings are particularly important for interpreting the results. First, similar functional performance does not imply similar static quality profiles across LLMs. Second, increases in raw static-analysis findings with task difficulty partly reflect increased implementation size and can weaken or reverse after length normalization. Third, functional correctness and static quality do not improve uniformly together: correct code may be more complete and structurally developed while still exhibiting greater complexity or different maintainability characteristics. Together, these findings show why correctness and static quality should be treated as complementary dimensions of LLM code-generation evaluation.

\subsection{Generalization beyond LiveCodeBench}

To assess the generalizability of our findings beyond LiveCodeBench, we repeated the same analysis with a subset of the LLMs used in this study on the HumanEval benchmark~\cite{chen2021evaluating}. From the 73 LLMs evaluated on LiveCodeBench, we selected those that were still available, including paid models with valid API access and open-source models with clearly specified parameter information. This resulted in a final subset of 29 LLMs for the analysis. Overall, the main trends remain broadly consistent, other than the relationship between correctness and structural code properties, which is comparatively weaker in HumanEval.

In LiveCodeBench, higher correctness is often associated with longer, more complex, and more explicitly structured code, reflected in higher LOC, LLOC, SLOC, Comments, Blank Lines, and Complexity. In HumanEval, the corresponding Radon metrics still vary across models, but the tendency for stronger or larger models to generate substantially more expanded code is less common. Instead, higher correctness in HumanEval is more often associated with fewer static-analysis findings, especially lower Pylint Avg Issues Per Code, Warning, and Error.

HumanEval also shows a more regular alignment between correctness and maintainability-oriented metrics. Higher-correctness models tend to exhibit higher Maintainability Index and lower SonarQube Maintainability Issues, whereas in LiveCodeBench the relationship between correctness and maintainability is less consistent. Taken together, these results suggest that several of our findings transfer beyond LiveCodeBench, but the strength and form of the observed quality profiles depend on the benchmark. LiveCodeBench more strongly links correctness with code expansion and structural complexity, while HumanEval more often aligns higher correctness with cleaner and more maintainable code.

{These benchmark-level differences indicate that static quality profiles are sensitive to task composition and benchmark characteristics. The supporting correct-only and length-normalized analyses further show that some structural differences persist after incomplete failed solutions are excluded, whereas some increases in raw issue counts are explained by differences in code volume. Accordingly, comparisons across benchmarks should distinguish between the overall issue burden of a generated solution and the concentration of findings relative to implementation size.}

\subsection{Implications}

Our findings suggest that pass-rate-based evaluation should be complemented with multidimensional static quality profiles. Metrics related to code size, complexity, maintainability, reliability, coding conventions, and security-related findings reveal model-specific tendencies that correctness alone does not capture. Correct-only and length-normalized analyses can further support interpretation by reducing the influence of incomplete implementations and code volume.

These findings have implications for several uses of LLM-generated code. Benchmark designers can complement pass rates with correctness-conditioned and length-normalized static quality metrics to avoid conflating functional success with code volume or implementation completeness. Model developers can use these profiles as secondary evaluation signals when models achieve similar functional performance. In practical development workflows, generated code that passes tests should still be subjected to static-analysis and quality checks before integration.

No single metric should be treated as a definitive measure of quality. Correctness, complexity, maintainability, and issue counts do not always vary in the same direction; therefore, static-analysis results should be interpreted jointly rather than collapsed into one score.

The observed positive associations between correctness and metrics such as code size, complexity, comments, and refactoring findings should also not be interpreted directly. They partly reflect that correct solutions are often more complete and structurally developed, creating more opportunities for static analyzers to report findings. These results may motivate prompt strategies that encourage complete and well-organized solutions, but they do not imply that longer or more complex code is inherently better.

Finally, these findings characterize function-level code generation in LiveCodeBench. Whether the same profiles hold for project-level, multi-file, and framework-dependent software remains an important direction for future work.

\subsection{Limitations and Future Directions}

The static quality profiles reported in this study should be interpreted in light of the benchmark, sampling procedure, and analysis tools used. LiveCodeBench consists mainly of short, function-level competitive programming tasks, which provide a controlled setting but cover only a limited range of software quality concerns. In particular, such tasks rarely include security-sensitive operations such as file I/O, shell commands, deserialization, authentication, or database access. As a result, the benchmark provides limited resolution for evaluating security. We observe the same pattern in HumanEval, where SonarQube security issues remain zero.

The correct-only analysis introduces an additional limitation because models with lower correctness contribute fewer programs, particularly on harder tasks. Consequently, some model--difficulty groups may be represented by substantially fewer samples than others, and their averages may be less stable. The correct-only results should therefore be interpreted together with the all-code results rather than as a replacement for them.

Length normalization also does not remove every source of variation. Metrics per 100 SLOC control for code volume, but they do not account for differences in algorithmic strategy, programming style, task-specific requirements, or the distribution of executable and non-executable lines. In addition, group-level normalization based on total findings divided by total SLOC gives greater weight to longer programs within each group. We use this formulation to estimate aggregate issue density, but alternative sample-level normalization strategies may produce different summaries.

In future work, we plan to evaluate static quality profiles on benchmarks requiring larger, more complex, and multi-file code generation, including tasks with greater coverage of security-relevant operations. Future studies could also evaluate whether the observed static-analysis findings correspond to developer judgments, maintenance effort, runtime failures, or vulnerabilities in project-level settings.

\subsection{Threats to Validity}

\textbf{Internal validity.}
Our results depend on the specific tools, versions, and configurations used for static analysis, including Pylint, Radon, and SonarQube. Static analysis is also inherently imperfect: tools may produce false positives, miss relevant issues, or reflect tool-specific rule definitions rather than actual runtime problems. We mitigate these threats by using stable tool settings and applying them consistently across all samples.

The extraction and filtering process may also affect the results. We select the first generated sample for each model--prompt pair and rely on the correctness labels reported by LiveCodeBench. Different sampling positions, decoding settings, or benchmark versions could produce different distributions of correctness and static quality metrics.

\textbf{Construct validity.}
The metrics used in this study are proxies for maintainability and reliability, and benchmark correctness is only an approximation of true functional correctness. We therefore interpret the observed relationships as descriptive associations rather than causal effects.

Raw issue counts and normalized issue densities represent different constructs. Raw counts measure the total findings present in a solution, whereas per-100-SLOC metrics measure findings relative to code volume. Neither should independently be treated as a complete measure of quality. Similarly, the correct-only subset controls for benchmark-level correctness but does not guarantee that passing programs are semantically correct for all possible inputs or suitable for production use.

\textbf{External validity.}
Our conclusions are limited to function-level benchmark settings such as LiveCodeBench and to Python code only, since all evaluated programs in this study are written in Python and the selected analysis tools are Python-focused. These tasks differ from real-world software systems, which are typically multi-file, evolve over time, and involve richer I/O and external dependencies. Some of the profiles observed here, especially the relationship between correctness, code size, and complexity, may therefore not transfer directly to project-level software settings or to code generation in other programming languages.

Although the HumanEval analysis provides an additional benchmark-level comparison, it also consists of short Python programming tasks and therefore does not fully address these external-validity limitations. Broader validation will require project-level benchmarks, additional programming languages, and code generated under realistic repository and maintenance contexts.

\section{Conclusion}

This paper presents a large-scale static quality profiling study of 65{,}663 Python programs generated by 73 LLMs. Using metrics from Pylint, Radon, and SonarQube, we characterize differences in coding conventions, complexity, size, maintainability, reliability, and security-related findings. Our results show that LLMs with similar functional performance can exhibit distinct static quality profiles. Maintainability and structural metrics provide the clearest differentiation, while reliability varies more moderately and security findings remain sparse. Harder tasks generally produce longer and more complex code, although changes in static-analysis findings differ across metrics. Correct-only and length-normalized analyses further show that several patterns persist after accounting for incomplete implementations and code volume.

These findings suggest that evaluations of LLM-generated code should complement correctness with multidimensional static quality profiles. Future work should extend this analysis to project-level, multi-file, and multi-language benchmarks.

\section{Acknowledgments}
This work was partially supported by The University of Queensland NSRSG Grant NS-2201 and Oracle Labs. 

%
%
\bibliographystyle{splncs04}
\bibliography{main}

@article{jiang2024survey,
  title={A survey on large language models for code generation},
  author={Jiang, Juyong and Wang, Fan and Shen, Jiasi and Kim, Sungju and Kim, Sunghun},
  journal={arXiv preprint arXiv:2406.00515},
  year={2024}
}

@article{nejjar2025llms,
  title={Llms for science: Usage for code generation and data analysis},
  author={Nejjar, Mohamed and Zacharias, Luca and Stiehle, Fabian and Weber, Ingo},
  journal={Journal of Software: Evolution and Process},
  volume={37},
  number={1},
  pages={e2723},
  year={2025},
  publisher={Wiley Online Library}
}

@article{liu2024exploring,
  title={Exploring and evaluating hallucinations in llm-powered code generation},
  author={Liu, Fang and Liu, Yang and Shi, Lin and Huang, Houkun and Wang, Ruifeng and Yang, Zhen and Zhang, Li and Li, Zhongqi and Ma, Yuchi},
  journal={arXiv preprint arXiv:2404.00971},
  year={2024}
}

@article{joel2024survey,
  title={A survey on llm-based code generation for low-resource and domain-specific programming languages},
  author={Joel, Sathvik and Wu, Jie and Fard, Fatemeh},
  journal={ACM Transactions on Software Engineering and Methodology},
  year={2024},
  publisher={ACM New York, NY}
}

@inproceedings{joshi2023repair,
  title={Repair is nearly generation: Multilingual program repair with llms},
  author={Joshi, Harshit and Sanchez, Jos{\'e} Cambronero and Gulwani, Sumit and Le, Vu and Verbruggen, Gust and Radi{\v{c}}ek, Ivan},
  booktitle={Proceedings of the AAAI Conference on Artificial Intelligence},
  volume={37},
  number={4},
  pages={5131--5140},
  year={2023}
}

@inproceedings{yang2024cref,
  title={Cref: An llm-based conversational software repair framework for programming tutors},
  author={Yang, Boyang and Tian, Haoye and Pian, Weiguo and Yu, Haoran and Wang, Haitao and Klein, Jacques and Bissyand{\'e}, Tegawend{\'e} F and Jin, Shunfu},
  booktitle={Proceedings of the 33rd ACM SIGSOFT International Symposium on Software Testing and Analysis},
  pages={882--894},
  year={2024}
}

@article{bouzenia2024repairagent,
  title={Repairagent: An autonomous, llm-based agent for program repair},
  author={Bouzenia, Islem and Devanbu, Premkumar and Pradel, Michael},
  journal={arXiv preprint arXiv:2403.17134},
  year={2024}
}

@inproceedings{pan2024lost,
  title={Lost in translation: A study of bugs introduced by large language models while translating code},
  author={Pan, Rangeet and Ibrahimzada, Ali Reza and Krishna, Rahul and Sankar, Divya and Wassi, Lambert Pouguem and Merler, Michele and Sobolev, Boris and Pavuluri, Raju and Sinha, Saurabh and Jabbarvand, Reyhaneh},
  booktitle={Proceedings of the IEEE/ACM 46th International Conference on Software Engineering},
  pages={1--13},
  year={2024}
}

@inproceedings{tao2024unraveling,
  title={Unraveling the Potential of Large Language Models in Code Translation: How Far are We?},
  author={Tao, Qingxiao and Yu, Tingrui and Gu, Xiaodong and Shen, Beijun},
  booktitle={2024 31st Asia-Pacific Software Engineering Conference (APSEC)},
  pages={353--362},
  year={2024},
  organization={IEEE}
}

@article{zhang2023planning,
  title={Planning with large language models for code generation},
  author={Zhang, Shun and Chen, Zhenfang and Shen, Yikang and Ding, Mingyu and Tenenbaum, Joshua B and Gan, Chuang},
  journal={arXiv preprint arXiv:2303.05510},
  year={2023}
}

@article{husein2025large,
  title={Large language models for code completion: A systematic literature review},
  author={Husein, Rasha Ahmad and Aburajouh, Hala and Catal, Cagatay},
  journal={Computer Standards \& Interfaces},
  volume={92},
  pages={103917},
  year={2025},
  publisher={Elsevier}
}

@inproceedings{cheng2025security,
  title={Security attacks on llm-based code completion tools},
  author={Cheng, Wen and Sun, Ke and Zhang, Xinyu and Wang, Wei},
  booktitle={Proceedings of the AAAI Conference on Artificial Intelligence},
  volume={39},
  number={22},
  pages={23669--23677},
  year={2025}
}

@phdthesis{rao2025navigating,
  title={Navigating challenges with LLM-based code generation using software-specific insights},
  author={Rao, Nikitha},
  year={2025},
  school={Carnegie Mellon University}
}

@article{mohsin2024can,
  title={Can we trust large language models generated code? a framework for in-context learning, security patterns, and code evaluations across diverse llms},
  author={Mohsin, Ahmad and Janicke, Helge and Wood, Adrian and Sarker, Iqbal H and Maglaras, Leandros and Janjua, Naeem},
  journal={arXiv preprint arXiv:2406.12513},
  year={2024}
}

@article{mu2024clarifygpt,
  title={Clarifygpt: A framework for enhancing llm-based code generation via requirements clarification},
  author={Mu, Fangwen and Shi, Lin and Wang, Song and Yu, Zhuohao and Zhang, Binquan and Wang, ChenXue and Liu, Shichao and Wang, Qing},
  journal={Proceedings of the ACM on Software Engineering},
  volume={1},
  number={FSE},
  pages={2332--2354},
  year={2024},
  publisher={ACM New York, NY, USA}
}

@article{dong2024self,
  title={Self-collaboration code generation via chatgpt},
  author={Dong, Yihong and Jiang, Xue and Jin, Zhi and Li, Ge},
  journal={ACM Transactions on Software Engineering and Methodology},
  volume={33},
  number={7},
  pages={1--38},
  year={2024},
  publisher={ACM New York, NY}
}

@article{gao2025current,
  title={The current challenges of software engineering in the era of large language models},
  author={Gao, Cuiyun and Hu, Xing and Gao, Shan and Xia, Xin and Jin, Zhi},
  journal={ACM Transactions on Software Engineering and Methodology},
  volume={34},
  number={5},
  pages={1--30},
  year={2025},
  publisher={ACM New York, NY}
}

@article{nijkamp2022codegen,
  title={Codegen: An open large language model for code with multi-turn program synthesis},
  author={Nijkamp, Erik and Pang, Bo and Hayashi, Hiroaki and Tu, Lifu and Wang, Huan and Zhou, Yingbo and Savarese, Silvio and Xiong, Caiming},
  journal={arXiv preprint arXiv:2203.13474},
  year={2022}
}

@article{dou2024s,
  title={What's wrong with your code generated by large language models? an extensive study},
  author={Dou, Shihan and Jia, Haoxiang and Wu, Shenxi and Zheng, Huiyuan and Zhou, Weikang and Wu, Muling and Chai, Mingxu and Fan, Jessica and Huang, Caishuang and Tao, Yunbo and others},
  journal={arXiv preprint arXiv:2407.06153},
  year={2024}
}

@inproceedings{zhong2024can,
  title={Can llm replace stack overflow? a study on robustness and reliability of large language model code generation},
  author={Zhong, Li and Wang, Zilong},
  booktitle={Proceedings of the AAAI conference on artificial intelligence},
  volume={38},
  number={19},
  pages={21841--21849},
  year={2024}
}

@inproceedings{chou2001empirical,
  title={An empirical study of operating systems errors},
  author={Chou, Andy and Yang, Junfeng and Chelf, Benjamin and Hallem, Seth and Engler, Dawson},
  booktitle={Proceedings of the eighteenth ACM symposium on Operating systems principles},
  pages={73--88},
  year={2001}
}

@inproceedings{palix2011faults,
  title={Faults in Linux: Ten years later},
  author={Palix, Nicolas and Thomas, Ga{\"e}l and Saha, Suman and Calv{\`e}s, Christophe and Lawall, Julia and Muller, Gilles},
  booktitle={Proceedings of the sixteenth international conference on Architectural support for programming languages and operating systems},
  pages={305--318},
  year={2011}
}

@inproceedings{brunet2009design,
  title={Design tests: An approach to programmatically check your code against design rules},
  author={Brunet, Joao and Guerrero, Dalton and Figueiredo, Jorge},
  booktitle={2009 31st International Conference on Software Engineering-Companion Volume},
  pages={255--258},
  year={2009},
  organization={IEEE}
}

@article{muccini2004using,
  title={Using software architecture for code testing},
  author={Muccini, Henry and Inverardi, Paola and Bertolino, Antonia},
  journal={IEEE Transactions on Software Engineering},
  volume={30},
  number={3},
  pages={160--171},
  year={2004},
  publisher={IEEE}
}

@book{aniche2022effective,
  title={Effective Software Testing: A developer's guide},
  author={Aniche, Maur{\'\i}cio},
  year={2022},
  publisher={Simon and Schuster}
}

@article{heumann2001generating,
  title={Generating test cases from use cases},
  author={Heumann, Jim},
  journal={The rational edge},
  volume={6},
  number={01},
  pages={1},
  year={2001}
}

@inproceedings{xia2024fuzz4all,
  title={Fuzz4all: Universal fuzzing with large language models},
  author={Xia, Chunqiu Steven and Paltenghi, Matteo and Le Tian, Jia and Pradel, Michael and Zhang, Lingming},
  booktitle={Proceedings of the IEEE/ACM 46th International Conference on Software Engineering},
  pages={1--13},
  year={2024}
}

@inproceedings{aschermann2019redqueen,
  title={REDQUEEN: Fuzzing with Input-to-State Correspondence.},
  author={Aschermann, Cornelius and Schumilo, Sergej and Blazytko, Tim and Gawlik, Robert and Holz, Thorsten},
  booktitle={NDSS},
  volume={19},
  pages={1--15},
  year={2019}
}

@misc{Pylint_contributors_Pylint,
  author = {{Pylint contributors}},
  title = {{Pylint}},
  howpublished = {\url{https://github.com/pylint-dev/pylint}},
  year = {2025}
}

@article{lacchia2023radon,
  title={radon-Code Metrics in Python},
  author={Lacchia, Michele and others},
  journal={URL https://github.com/rubik/radon},
  year={2023}
}

@book{papapetrou2013sonarqube,
  title={SonarQube in action},
  author={Papapetrou, Patroklos},
  year={2013},
  publisher={Simon and Schuster}
}

@inproceedings{krishnamoorthy2010tackling,
  title={Tackling the path explosion problem in symbolic execution-driven test generation for programs},
  author={Krishnamoorthy, Saparya and Hsiao, Michael S and Lingappan, Loganathan},
  booktitle={2010 19th IEEE Asian Test Symposium},
  pages={59--64},
  year={2010},
  organization={IEEE}
}

@book{copeland2005pmd,
  author    = {Tom Copeland},
  title     = {PMD Applied},
  year      = {2005},
  publisher = {Centennial Books},
}

@article{baggen2012standardized,
  title={Standardized code quality benchmarking for improving software maintainability},
  author={Baggen, Robert and Correia, Jos{\'e} Pedro and Schill, Katrin and Visser, Joost},
  journal={Software Quality Journal},
  volume={20},
  number={2},
  pages={287--307},
  year={2012},
  publisher={Springer}
}

@article{syed2015guidelines,
  title={Guidelines for establishing reliability when coding narrative data},
  author={Syed, Moin and Nelson, Sarah C},
  journal={Emerging Adulthood},
  volume={3},
  number={6},
  pages={375--387},
  year={2015},
  publisher={Sage Publications Sage CA: Los Angeles, CA}
}

@inproceedings{braz2022software,
  title={Software security during modern code review: the developer’s perspective},
  author={Braz, Larissa and Bacchelli, Alberto},
  booktitle={Proceedings of the 30th ACM Joint European Software Engineering Conference and Symposium on the Foundations of Software Engineering},
  pages={810--821},
  year={2022}
}

@article{chen2021evaluating,
  title={Evaluating large language models trained on code},
  author={Chen, Mark},
  journal={arXiv preprint arXiv:2107.03374},
  year={2021}
}

@article{austin2021program,
  title={Program synthesis with large language models},
  author={Austin, Jacob and Odena, Augustus and Nye, Maxwell and Bosma, Maarten and Michalewski, Henryk and Dohan, David and Jiang, Ellen and Cai, Carrie and Terry, Michael and Le, Quoc and others},
  journal={arXiv preprint arXiv:2108.07732},
  year={2021}
}

@inproceedings{lai2023ds,
  title={DS-1000: A natural and reliable benchmark for data science code generation},
  author={Lai, Yuhang and Li, Chengxi and Wang, Yiming and Zhang, Tianyi and Zhong, Ruiqi and Zettlemoyer, Luke and Yih, Wen-tau and Fried, Daniel and Wang, Sida and Yu, Tao},
  booktitle={International Conference on Machine Learning},
  pages={18319--18345},
  year={2023},
  organization={PMLR}
}

@article{liu2023your,
  title={Is your code generated by chatgpt really correct? rigorous evaluation of large language models for code generation},
  author={Liu, Jiawei and Xia, Chunqiu Steven and Wang, Yuyao and Zhang, Lingming},
  journal={Advances in Neural Information Processing Systems},
  volume={36},
  pages={21558--21572},
  year={2023}
}

@article{jain2024livecodebench,
  title={Livecodebench: Holistic and contamination free evaluation of large language models for code},
  author={Jain, Naman and Han, King and Gu, Alex and Li, Wen-Ding and Yan, Fanjia and Zhang, Tianjun and Wang, Sida and Solar-Lezama, Armando and Sen, Koushik and Stoica, Ion},
  journal={arXiv preprint arXiv:2403.07974},
  year={2024}
}

@article{tornhill2022code,
  title={Code Red: The business impact of code quality},
  author={Tornhill, A and Borg, M},
  journal={arXiv preprint arXiv:2203.04374},
  year={2022}
}

@misc{Sonar2023,
  author       = "{SonarSource}",
  title        = "{The State of Technical Debt 2023}",
  year         = {2023},
  howpublished = {\url{https://www.sonarsource.com}},
}

@misc{McKinsey2023,
  author       = "{McKinsey \& Company}",
  title        = "{Tech Debt: The Hidden Cost of Software Complexity}",
  year         = {2023},
  howpublished = {\url{https://www.mckinsey.com}},
}

@inproceedings{gulabovska2019survey,
  title={Survey on Static Analysis Tools of Python Programs.},
  author={Gulabovska, Hristina and Porkol{\'a}b, Zolt{\'a}n},
  booktitle={SQAMIA},
  year={2019}
}

@article{hassan2024evaluating,
  title={Evaluating python static code analysis tools using FAIR principles},
  author={Hassan, Hassan Bapeer and Sarhan, Qusay Idrees and Besz{\'e}des, {\'A}rp{\'a}d},
  journal={IEEE Access},
  volume={12},
  pages={173647--173659},
  year={2024},
  publisher={IEEE}
}

@article{lenarduzzi2020some,
  title={Some sonarqube issues have a significant but small effect on faults and changes. a large-scale empirical study},
  author={Lenarduzzi, Valentina and Saarim{\"a}ki, Nyyti and Taibi, Davide},
  journal={Journal of Systems and Software},
  volume={170},
  pages={110750},
  year={2020},
  publisher={Elsevier}
}

@INPROCEEDINGS{11126649,
  author={Sepidband, Melika and Taherkhani, Hamed and Wang, Song and Hemmati, Hadi},
  booktitle={2025 IEEE 49th Annual Computers, Software, and Applications Conference (COMPSAC)}, 
  title={Enhancing LLM-Based Code Generation with Complexity Metrics: A Feedback-Driven Approach}, 
  year={2025},
  volume={},
  number={},
  pages={1416-1426},
  doi={10.1109/COMPSAC65507.2025.00178}}

@INPROCEEDINGS{11185875,
  author={Afrin, Saima and Xu, Bowen and Mastropaolo, Antonio},
  booktitle={2025 IEEE International Conference on Software Maintenance and Evolution (ICSME)}, 
  title={Is Quantization a Deal-Breaker? Empirical Insights From Large Code Models}, 
  year={2025},
  volume={},
  number={},
  pages={1-13},
  doi={10.1109/ICSME64153.2025.00049}}

@article{idrisov2024program,
  title={Program code generation with generative AIs},
  author={Idrisov, Baskhad and Schlippe, Tim},
  journal={Algorithms},
  volume={17},
  number={2},
  pages={62},
  year={2024},
  publisher={MDPI}
}
\end{document}